\documentclass[aps,prb,twocolumn,superscriptaddress,showpacs]{revtex4-2}
\usepackage{graphicx,bm,color,float,textcomp,mathtools,times,nicefrac}
\newcommand\sho{$\rm SrHo_2O_4$}
\newcommand\sgo{$\rm SrGd_2O_4$}
\newcommand\seo{$\rm SrEr_2O_4$}
\newcommand\sdo{$\rm SrDy_2O_4$}
\newcommand\syo{$\rm SrYb_2O_4$}
\newcommand\sno{$\rm SrNd_2O_4$}
\newcommand\sro{SrRE$_2$O$_4$}
\newcommand\bto{$\rm BaTb_2O_4$}
\newcommand\bdo{$\rm BaDy_2O_4$}
\newcommand\sto{$\rm SrTb_2O_4$}
\newcommand\stmo{$\rm SrTm_2O_4$}
\newcommand\tb{Tb$^{3+}$}
\newcommand\afm{antiferromagnetic}
\newcommand\Tn{$T_{\rm N}$}

\newcommand\cef{CEF}

\usepackage{soul,xcolor}
\usepackage{ulem}
\usepackage{hyperref}
\setstcolor{magenta}

\begin{document}
\title{Magnetic properties of the zigzag ladder compound SrTb$_2$O$_4$}
\date{\today}
	\author{F. Orlandi}		\affiliation{ISIS Neutron and Muon Source, STFC Rutherford Appleton Laboratory, Chilton, Didcot, OX11 0QX, United Kingdom}
	\author{M.~Ciomaga~Hatnean} \altaffiliation[Current addresses: ]{Laboratory for Multiscale materials eXperiments, Paul Scherrer Institute, 5232 Villigen PSI, Switzerland \& Materials Discovery Laboratory, Department of Materials, ETH Zurich, 8093 Zurich, Switzerland}	
	\affiliation{Department of Physics, University of Warwick, Coventry, CV4 7AL, United Kingdom}
	\author{D.A.~Mayoh}			\affiliation{Department of Physics, University of Warwick, Coventry, CV4 7AL, United Kingdom}
	\author{J.P.~Tidey}			\affiliation{Department of Chemistry, University of Warwick, Coventry, CV4 7AL, UK}
	\author{S.X.M.~Riberolles}	\affiliation{Department of Physics, University of Warwick, Coventry, CV4 7AL, United Kingdom}
	\author{G.~Balakrishnan}		\affiliation{Department of Physics, University of Warwick, Coventry, CV4 7AL, United Kingdom}											
	\author{P.~Manuel}			\affiliation{ISIS Neutron and Muon Source, STFC Rutherford Appleton Laboratory, Chilton, Didcot, OX11 0QX, United Kingdom}
	\author{D.D.~Khalyavin}		\affiliation{ISIS Neutron and Muon Source, STFC Rutherford Appleton Laboratory, Chilton, Didcot, OX11 0QX, United Kingdom}
	\author{H.C.~Walker}		\affiliation{ISIS Neutron and Muon Source, STFC Rutherford Appleton Laboratory, Chilton, Didcot, OX11 0QX, United Kingdom}
	\author{M.D.~Le}			\affiliation{ISIS Neutron and Muon Source, STFC Rutherford Appleton Laboratory, Chilton, Didcot, OX11 0QX, United Kingdom}													
	\author{B.~Ouladdiaf}		\affiliation{Institut Laue-Langevin, 71 Avenue des Martyrs, CS 20156, 38042 Grenoble Cedex 9, France}
	\author{A. R. Wildes}			\affiliation{Institut Laue-Langevin, 71 Avenue des Martyrs, CS 20156, 38042 Grenoble Cedex 9, France}
	\author{N.~Qureshi}			\affiliation{Institut Laue-Langevin, 71 Avenue des Martyrs, CS 20156, 38042 Grenoble Cedex 9, France}			
	\author{O.A.~Petrenko}		\email{O.Petrenko@warwick.ac.uk} \affiliation{Department of Physics, University of Warwick, Coventry, CV4 7AL, United Kingdom}
		
\begin{abstract}	
We report on the properties of \sto, a frustrated zigzag ladder antiferromagnet, studied by single crystal neutron diffraction (with polarised neutrons in zero field and unpolarised neutrons in an applied magnetic field), as well as by neutron spectroscopy on a polycrystalline sample.
The neutron scattering results are supported by single crystal magnetisation and heat capacity measurements.
In zero field, neutron diffraction data show no transition to a magnetically ordered state down to the lowest experimentally available temperature of 35~mK, and the material remains magnetically disordered down to this temperature.
Polarised neutron diffraction measurements reveal the presence of a diffuse scattering signal suggesting only very weak spin-spin correlations in the ground state.
For $H\! \parallel \! c$ (the easy magnetisation direction), we followed the magnetisation process using neutron diffraction measurements and observed the appearance of field-induced magnetic Bragg peaks with integer $h$ and $k$ indices in the $(hk0)$ scattering plane.
No magnetic peaks with a non-zero propagation vector were detected.
The observed in-field data fit well to a simple two-sublattice model with magnetic moments aligned along the field direction but being significantly different in magnitude for the two inequivalent \tb\ sites in the unit cell.
Overall, the collected data point to a nonmagnetic ground state in \sto\ despite the presence of strong interactions.
\end{abstract}

\maketitle
\section{Introduction}
\begin{figure}[tb]
\includegraphics[width=0.99\columnwidth]{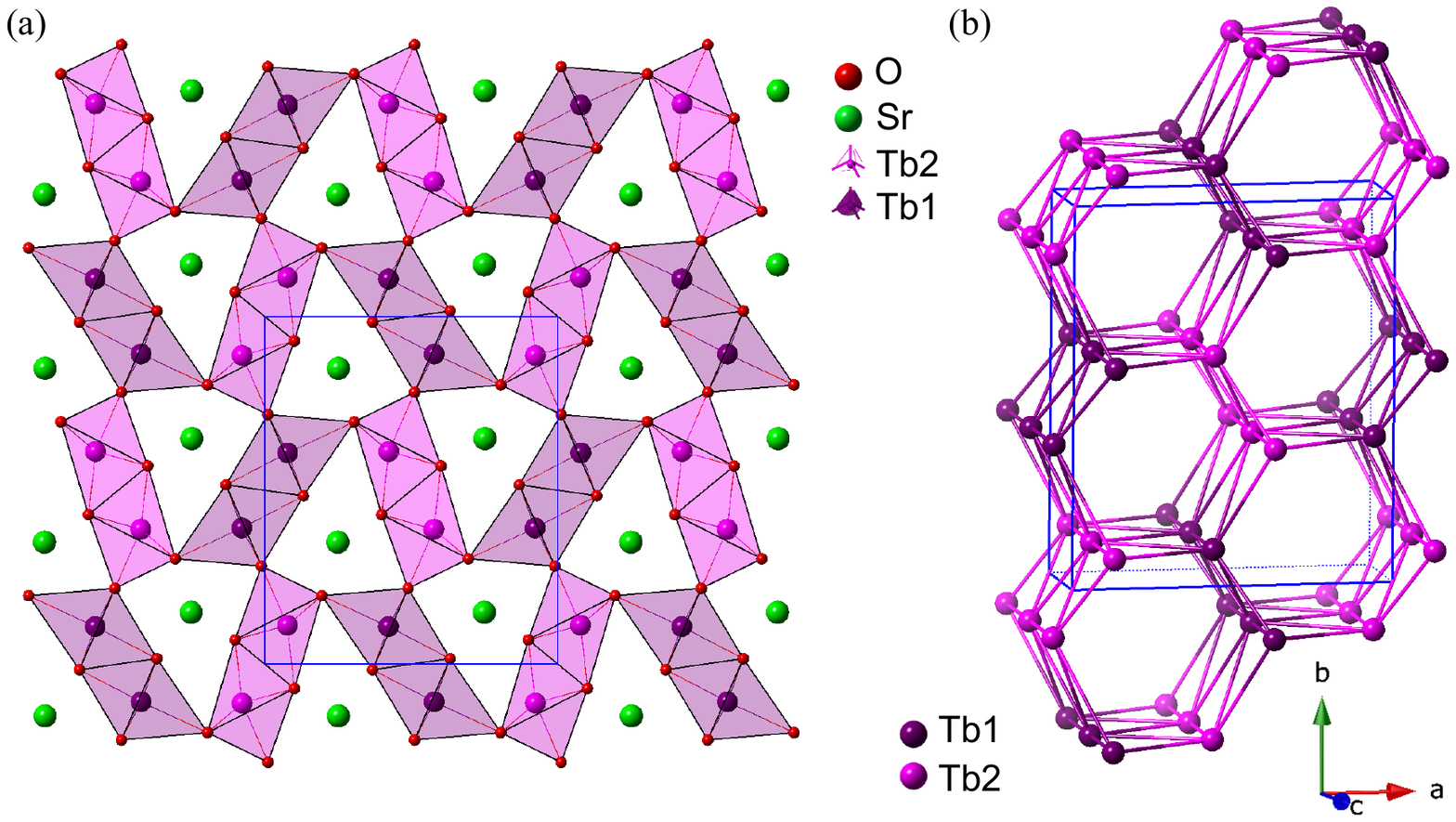}
\vspace{-5mm}
\caption{(a) Crystal structure of \sto\ projected onto the $a$-$b$ plane.
Different colours used for the oxygen octahedra hosting the Tb ions emphasise the presence of two inequivalent crystallographic positions, labeled as Tb1 and Tb2.
(b) The zigzag ladders formed by the Tb1 and Tb2 magnetic ions and their arrangement into a distorted honeycomb lattice in \sto.
The unit cell is indicate by the solid blue lines for panels (a) and (b).}
\label{fig:1_structure}
\end{figure}

\sto\ is a member of a large family of the \sro\ oxides, where RE is a rare-earth atom~\cite{Karunadasa_2005,Fennell_2014,Wen_2015}.
The \sro\ magnets demonstrate rather interesting properties mostly originating from their crystal structure (space group $Pnam$) which for magnetic ions consists of triangular zigzag ladders, as depicted in Fig.~\ref{fig:1_structure}.
Due to the resulting geometrical frustration, the ordering temperatures \Tn\ for most family members are significantly lower than their Weiss temperatures, $\Theta_{\rm CW}$.
In many cases the magnetic order is incomplete, and the range of the spin correlations remains limited even at the lowest temperatures, although it might extend over several unit cell lengths.
Another factor influencing the unconventional behaviour of the \sro\ magnets is the presence of two different crystallographic positions for the RE ions in the orthorhombic unit cell (in \sto, they are labelled as Tb1 and Tb2 in  Fig.~\ref{fig:1_structure}).
Despite rather similar crystallographic arrangements for the RE ions in the two positions, with only marginally more distorted oxygen octahedra surrounding one of them, they usually demonstrate very different magnetic behaviour~\cite{Hayes_2011,Young_2013,Bidaud_2016,Qureshi_2021_b}.  

The magnetic properties of the RE containing materials are often governed by spin anisotropy which in turn is determined by crystal electric field (\cef) effects.
Ising-like, XY-like and Heisenberg-like behaviours are all found in different \sro\ magnets.
When studying the families of the RE containing compounds which share the same chemical composition and crystal symmetry, but contain different RE elements, a better understanding of the magnetic properties of the entire family can be achieved by comparing and contrasting the \cef\ level schemes of the individual compounds.
The behaviour of the magnets containing \tb\ ions ($S=L=3$, $J=6$) is often special in this respect, as singlet or nonmagnetic ground states are frequently found.
This possibility makes studies of the \tb\ containing materials both challenging and exciting, Tb$_2$Ti$_2$O$_7$ is a good example of a magnetic system which required more than two decades of concerted efforts~\cite{Gardner_1999,Gingras_2000,Ruminy_2019,Slobinsky_2021} to tackle what is considered to be an ``ongoing conundrum'' in the study of RE pyrochlores.

To properly place \sto\ among other \sro\ family members, we briefly list here their magnetic properties. 
\seo\ undergoes a transition at 0.75~K~\cite{Petrenko_2008} with magnetic moments on the Er1 site forming a fully ordered ${\bf q}=0$ antiferromagnet, while the magnetic moments on the Er2 site form a collection of weakly-correlated \afm\ chains with only a short-range order described by a ${\bf q}=(00q_z)$ propagation vector with $q_z \approx 0.5$~\cite{Hayes_2011}.
A similar coexistence of two types of magnetic order is also found in \sho, but in this compound even the ${\bf q}=0$ structure remains short-ranged down to the lowest temperatures despite a pronounced increase in the magnetic correlations below 0.7~K~\cite{Young_2013}.
Long-range order in \sdo\ is absent down to the mK range~\cite{Petrenko_2017,Gauthier_2017a,Gauthier_2017b}. 
\syo\ orders at 0.9~K to a noncollinear structure with a reduced moment on both Yb sites~\cite{Quintero_2012}.
\sno\ orders at 2.28~K with only one Nd site carrying a significant magnetic moment~\cite{Qureshi_2021_a}.
\sgo\ shows an initial transition at 2.73~K to a commensurate \afm\ structure and then a further transition at 0.48~K to an incommensurate structure with ${\bf q}=(0~0~0.42)$ where both Gd sites are long-range ordered~\cite{Qureshi_2022}. 

In an applied magnetic field, all the \sro\ compounds demonstrate highly anisotropic behaviour often forming rather complex, partially ordered states.
The in-field ground-states of \seo, \sho, \sdo, and perhaps other \sro\ family members could adequately be described by an Ising model on a honeycomb zigzag-ladder lattice with two different types of magnetic sites~\cite{Dublenych_2022}.

Given the above, the reported magnetic properties of \sto~\cite{Li_2014_STO}  are surprising from at least two perspectives.
The first one is the highest ordering temperature of 4.28~K~\cite{Li_2014_STO} in the whole family. 
The second is the observed incommensurate magnetic structure with a propagation vector of ${\bf q}\approx(0.5924~0.0059~0)$~\cite{Li_2014_STO}, implying ferromagnetic ordering along the $c$~axis and an incommensurate modulation along the $a$ and $b$~axes.
It is difficult to understand what combination of the magnetic coupling constants, either of exchange or the dipole-dipole origin, could potentially cause such an arrangement.
None of the other family members have a similar propagation vector.
For all of the studied \sro\ compounds, the $a^\ast$ component of the propagation vector, $q_x$, is zero.
In fact, the only example of the ordering with a non-zero $q_x$ is reported for \bdo~\cite{Prevost_2018}, a member of the BaRE$_2$O$_4$ family with an identical crystal structure~\cite{Doi_2006}.
However, in \bdo\ the two propagation vectors, ${\bf q}_1=(\frac{1}{2}0\frac{1}{2})$ and ${\bf q}_2=(\frac{1}{2}\frac{1}{2}\frac{1}{2})$ both have non-zero $c^\ast$ components, $q_z$, indicating an \afm\ arrangement along the $c$~axis, which makes the magnetic structure much more ``rational'' -- an increase of the unit cell along only the $a$~axis, without altering the arrangements within the zigzag ladders running along the $c$~axis is unlikely to relieve the geometrical frustration.

Notably, the \tb\ compound \bto\ from the same BaRE$_2$O$_4$ family is found to be in a cooperative paramagnetic or spin-liquid ground state down to 95~mK~\cite{Aczel_2015}.

In this paper we report the preparation of high quality single crystal samples of \sto\ and a detailed investigation of their magnetic properties. 
In zero field, the samples remain in a magnetically disordered state down to at least 35~mK.
The whole volume of the collected data is consistent with a picture of a magnetically disordered ground state and a field-induced magnetic polarisation of the magnetic \tb\ ions.

\section{Samples preparation and experimental details} \label{sec:experimental}
A polycrystalline sample of \sto\ was prepared by grinding together stoichiometric SrCO$_3$ (99.999\%, Strem Chemicals) and Tb$_4$O$_7$ (99.9\%, Strem Chemicals) with an excess of 12.5\% SrCO$_3$ to compensate for the loss of Sr during the synthesis.
The resulting ground powder was pressed into pellets of approximately 2~g each.
The pellets were then sintered in an argon-hydrogen (Ar + 2\% H$_2$) atmosphere at 900~$^\circ$C and 1450~$^\circ$C for 12 and 48 hours respectively, with intermediate regrinding and re-pressing.
The samples prepared contained a few percent of Tb$_2$O$_3$ and SrO$_2$ impurities.

A single crystal of \sto\ was grown using the optical floating-zone technique (an image of the crystal boule is given in Fig.~S1 in the Supplemental Material~\cite{Supp_2025}).
Polycrystalline rods were used for both the feed and seed rods for the crystal growth.
The crystal growth was carried out in a high purity (99.9999\%) argon gas pressure of about 1~bar and at a growth rate of 3-5~mm/h.

Powder x-ray diffraction was performed using Panalytical X-Pert Pro MPD which has a Cu x-ray source that is equipped with a focusing Johanson monochromator to the incident optics to give monochromatic K$_{\alpha1}$ radiation ($\lambda = 1.5406$~\AA).

Single crystal x-ray diffraction data for \sto\ were collected for a crystal of dimension $0.04 \times 0.06 \times 0.13$~mm$^3$ using a Rigaku Oxford Diffraction Synergy S equipped with a HyPix-6000HE Hybrid Photon Counting detector and employing mirror monochromated Mo K$_{\alpha}$ radiation ($\lambda = 0.71073$~\AA) source.
The temperature of the crystal was controlled using an Oxford Cryosystems Cryostream at 300~K.
CrysAlisPRO was used for the data collection, indexing, reduction and absorption corrections~\cite{Rigaku19}.
A structural solution was obtained using SHELXT~\cite{Sheldrick15_XT} and further refined by full-matrix least squares using SHELXL~\cite{Sheldrick15_XL}, both operating through OLEX2~\cite{Dolomanov09}.

High crystallinity of the grown crystal was confirmed by x-ray and neutron Laue diffraction techniques.
A single crystal sample, $m \! \approx \! 120$~mg, was used for the zero-field Laue diffraction measurements on CYCLOPS~\cite{CYCLOPS} and polarised neutron diffraction measurements on D7~\cite{Stewart_2009} at the Institut Laue-Langevin (Grenoble, France) as well as for the in-field measurements on WISH time-of-flight instrument~\cite{WISH_2011} at ISIS (STFC Rutherford Appleton Laboratory, United Kingdom).

CYCLOPS is a very large solid-angle Laue neutron diffractometer using CCD detector with a very fast read-out system allowing real-time Laue diffraction experiments~\cite{CYCLOPS}.
An orange cryostat has been used to cover the temperature range 2 to 10~K in the experiment.
A set of Laue patterns has been collected at different omega angles to access a large volume of reciprocal space especially at low scattering angles relevant for magnetic reflections.

The magnetic correlations were probed using D7, the diffuse scattering spectrometer with polarization analysis ~\cite{D7_2018}.  The single crystal sample was mounted on an aluminium pin, and the pin was masked with cadmium.
The horizontal $(hk0)$ plane was explored by rotating the sample around the vertical $c$~axis in $1^\circ$ steps, two different positions of the detectors allowed for a denser coverage of the reciprocal space.
The incident neutrons wavelength was $\lambda = 3.1$~\AA.
During the experiment, the neutron polarisation was normal to the scattering plane and non-spin-flip (NSF) and spin-flip (SF) datasets were collected.  A standard orange cryostat provided a base temperature of 1.5~K, and the instrument background was estimated by measuring the empty cryostat.

Unpolarized single crystal measurements were performed as a function of applied magnetic field on WISH~\cite{WISH_2019}.  A dilution refrigerator with base temperature of 35~mK was used in combination with a 10~T cryomagnet.
An oxygen-free copper sample holder provided good thermal link from the mixing chamber to the sample.
The vertical magnetic field was applied along the $c$~axis, this setup provided access to the horizontal $(hk0)$ scattering plane.
We used three different sample positions to maximise neutron flux for different portions of reciprocal space and to collect the intensity of a large number of magnetic and nuclear peaks.
Typical measurement time for a single position was 30~minutes.
The integration of the diffracted peaks at 0~kOe and 80~kOe was performed with the Mantid software~\cite{Arnold_2014}.
The data were corrected for incident flux, Lorentz factor and detector efficiency.
The integration has been performed in $Q$-space using an elliptical region of interest on predicted peaks position.
The main axes of the ellipse for each reflection were determined from the estimated covariance of the data.
The refinements were performed with the help of the Jana2006 software~\cite{Petvrivcek_2014} and group theoretical calculations were performed with the ISOTROPY suite of software~\cite{Stokes_2006}.

Inelastic neutron scattering experiments were performed using 2.6~g of polycrystalline \sto\ on the MERLIN direct geometry chopper spectrometer~\cite{Bewley_2009} at the ISIS neutron and muon facility.
The sample was wrapped in aluminium foil to form an annulus and inserted into a closed cycle refrigerator in a cylindrical aluminium can.
Two chopper configurations were used to give focused incident energies of 7 and 30~meV at frequencies of 150 and 300~Hz respectively.
Each configuration also yielded additional ``reps'' at 18~meV (150~Hz setting) and 80~meV (300~Hz setting) using repetition-rate multiplication method.
The raw data~\cite{MERLIN_2019} were processed using the Mantid software package~\cite{Arnold_2014} following the standard conventions.

Magnetisation measurements were performed on both single crystal and polycrystalline samples of \sto\ using a Quantum Design Magnetic Property Measurement System (MPMS) magnetometer as well as an Oxford Instruments vibrating sample magnetometer (VSM).
Specific heat was measured on single crystal samples by the standard  2$\tau$ method using a Quantum Design Physical Property Measurement System (PPMS) calorimeter in the temperature range 1.8 -- 50~K in applied fields of 0, 20, 40, 60, and 80~kOe.
The zero-field data collection was extended down to 0.6~K by using a $^3$He option.

\section{Results, analysis and discussion}
\subsection{Powder and single crystal x-ray diffraction}
\begin{figure}[tb]
\includegraphics[width=0.99\columnwidth]{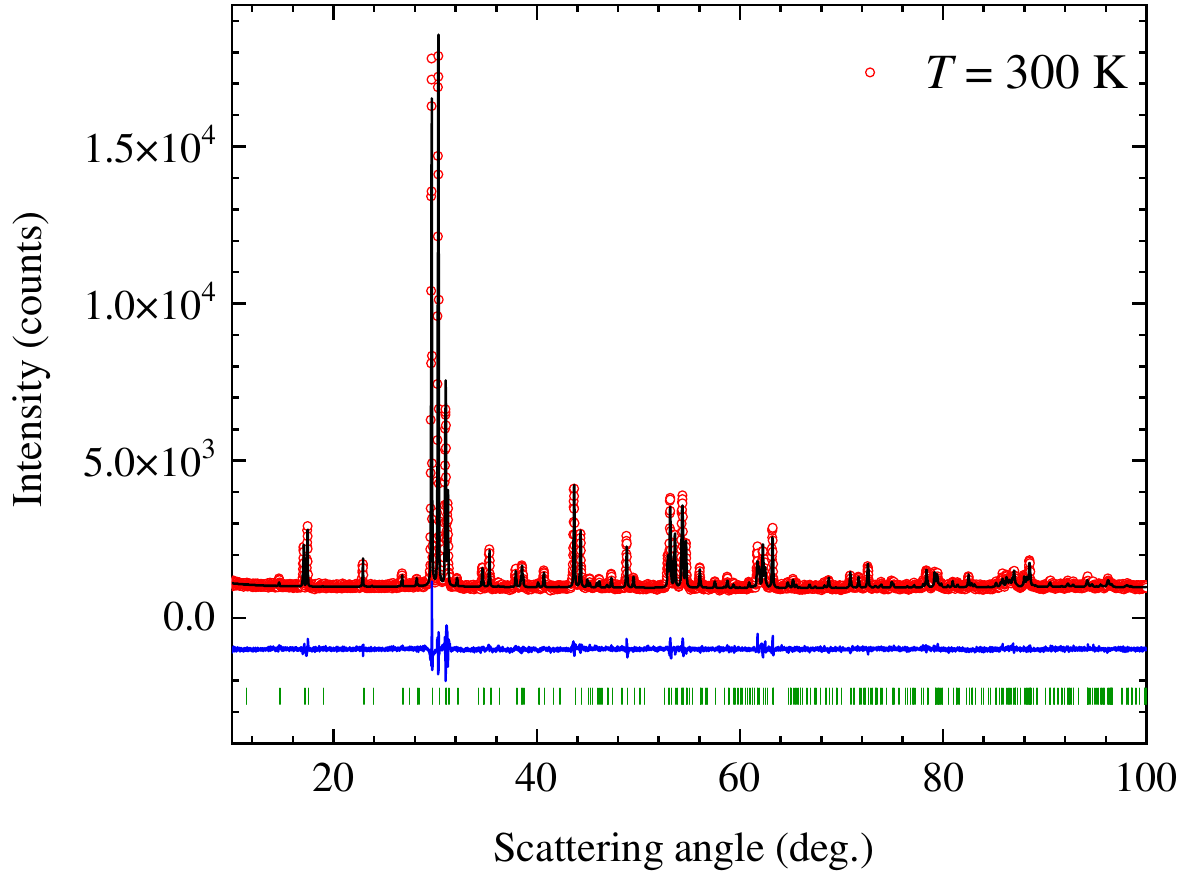}
\caption{Refinement of the powder x-ray diffraction profile of the ground crystal of \sto\ measured at room temperature using a Cu x-ray source. 
Data, fit, and their difference are displayed in red, black, and blue, respectively.
The positions of the nuclear reflections allowed within the $Pnam$ crystal structure of \sto\ are indicated by vertical green bars.}
\label{fig:2_XRD}
\end{figure}

Powder x-ray diffraction patterns taken on ground crystals of \sto\ revealed their high chemical purity (see Fig.~\ref{fig:2_XRD}), with no detectable impurity peaks.
Therefore, the powder sample used on MERLIN was prepared by grinding the grown crystals.
A record of the crystal structure parameters determined from XRD patterns refinement is summarised in Table~\ref{tab:Xrays}.
The crystal structure parameters are similar to those reported in Ref.~\cite{Li_2014_STO}.

\begin{table}[b]
\caption{Crystal data, reliability factors and structural parameters resulting from the single crystal and powder x-ray diffraction experiments.
The space group is $Pnam$, the atomic positions are all at the 4c Wyckoff site (x,y,1/4).
For the PXRD, the isotropic atomic displacement parameter $B_{\rm iso}$ was fixed to 0.97~\AA$^2$ for all atoms, whereas for the single crystal data atomic displacement parameters were refined independently for each site and are reported in the cif file provided as Supplemental Material~\cite{Supp_2025}.}
\label{tab:Xrays}
\begin{tabular}{c|c|c}
\hline 
\hline  
Technique			& Single-crystal XRD		& PXRD			     \\	\hline  
a (\AA)				& 10.0827(2) 				& 10.10385(8)		 \\
b (\AA)				& 11.9876(4) 				& 12.00133(9)  		 \\
c (\AA)				& 3.4506(1)     			& 3.45324(4)		 \\
$R_1$ (\%)			& 1.79						&				     \\
$wR_{2}$ (\%)		& 4.4						&				     \\
$R_{wp}$ 			&							& 4.474			     \\
$R_{exp}$			&							& 3.307              \\
$R_{Bragg}$			&							& 1.189			     \\
$GoF$ (\%)			& 1.325						&1.496			     \\
Sr x,  y			& ~~0.75073(3) 0.64866(3)	& ~~0.7510(4) 0.6492(3)\\
O1 x, y				& 0.2142(3) ~0.1801(2) 		& ~~0.220(3) ~0.1822(16)\\
O2 x, y				& 0.1291(3)  ~0.4825(2)		& ~~0.127(2) ~0.4926(19)\\
O3 x, y				& 0.5099(3)  ~0.7846(2) 	& ~~0.481(3) ~0.7792(15)\\
O4 x, y				& 0.4265(3)  ~0.4214(2)		& ~~0.418(4) ~0.4064(16)\\
Tb1 x, y			& ~~0.42484(2)  0.11191(2) 	& ~~0.4232(4) ~0.1122(2)\\
Tb2 x, y			& ~~0.41883(2)  0.61157(2) 	& ~~0.4199(4) ~0.6111(2)\\
\hline                      
\hline
 \end{tabular}
\end{table}

Single crystal x-ray diffraction was used to further confirm the structure of the single crystals of \sto\ grown by the floating zone technique.
The structural parameters derived from the single crystal x-ray diffraction data are listed in Table~\ref{tab:Xrays}, and a cif file containing all the information is provided as Supplemental Material~\cite{Supp_2025}.
It is also worth underlining good agreement between the powder and single crystal refinement indicating high sample quality and the lack of strain or inhomogeneities in the samples.

\subsection{Inelastic neutron scattering}  \label{sec:INS}

\begin{figure}[tb]
\includegraphics[width=0.99\columnwidth]{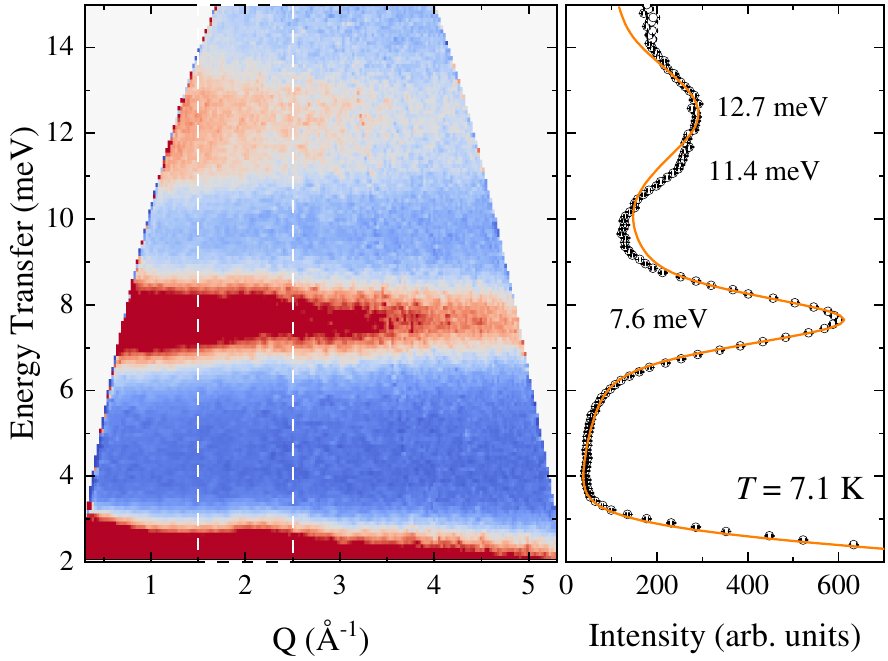}
\includegraphics[width=0.99\columnwidth]{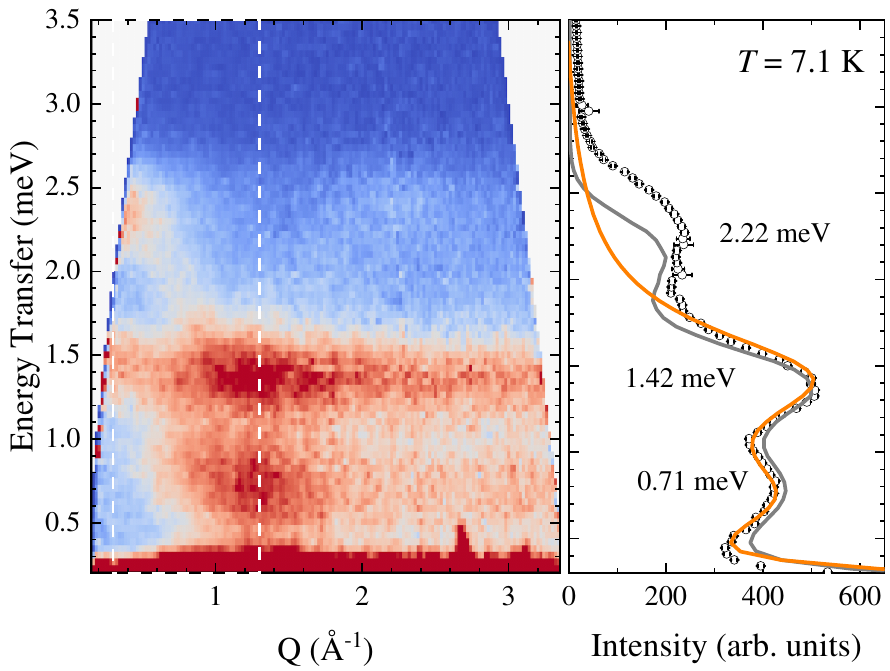}
\caption{INS spectra of \sto\ powder sample measured on MERLIN at $T = 7.1$~K with different energy neutrons, (top) $E_i=18$~meV and (bottom) $E_i=7$~meV.
To identify the exact energies of the observed excitations, the cuts along the energy transfer are shown to the right of the colour intensity maps.
White dashed lines on the colour maps indicate the area used for the cuts.
The experimental data (black symbols) were fitted as Gaussians and a linear background, with the peak centres shown near each peak.
The calculated spectra from parameters in Table~\ref{cef_pars} are shown in both panels by the solid orange lines.
In the lower panel, the grey line shows the calculated spectrum using the mean-field random-phase-approximation.}
\label{fig:3_MERLIN}
\end{figure} 

Inelastic neutron scattering (INS) spectra measured on the time-of-flight MERLIN spectrometer at ISIS are shown in Fig.~\ref{fig:3_MERLIN}. Spectra with incident neutrons of 18 and 30~meV were found to be very similar so for brevity we show only the 18~meV data.
The latter data, measured at $T=7$~K, revealed the presence of three well-defined \cef\ excitations at 7.6,  11.4, and 12.7~meV (the higher energy branches are slightly overlapping).
The excitations are largely non-dispersive, although there is a visible modulation of the intensity of the 7.6~meV branch with increasing scattering vector $Q$.

The lower-energy excitations observed with incident neutrons of 7~meV demonstrate a significantly more complex behaviour.
There are two strong, nearly flat excitation branches at 0.71 and 1.42~meV whose intensity is maximum around $Q \approx 1.25$~\AA$^{-1}$. 
The third, much weaker excitation branch at 2.22~meV appears to be strongly dispersive with an energy minimum at the same $Q$.

The measurements with the highest energy neutrons of 82~meV (shown in Fig.~S2 in the Supplemental Material~\cite{Supp_2025}) revealed the presence of two further partially overlapping \cef\ excitations at 27.3 and 31.1~meV.
There are no visible \cef\ levels above these energies all the way up to at least 70~meV, with the excitation spectrum at higher scattering vector, $Q>6$~\AA$^{-1}$ being dominated by the phonons.

The temperature evolution of the \cef\ energy spectrum is shown in Fig.~S3 in the Supplemental Material~\cite{Supp_2025}.

These results indicate a potential influence of exchange interactions on the CEF excitations, although the strength of the exchange coupling appears to be insufficient to drive the system into a magnetically ordered state, so it remains in a disordered (correlated paramagnetic) phase~\cite{Wang_1968,Jensen_1991}. 

In addition to being responsible for the dispersion of the 2.22~meV mode, the exchange interaction broadens the linewidths of the CEF excitations and makes it more difficult to distinguish if there are multiple nearby peaks.
Furthermore, while the RE sites in the \sro\ compounds have monoclinic point symmetry $C_s$($m$), which leads to the splitting into the maximum number of the levels, in the case of non-Kramer's \tb\ to $2J+1=13$ singlets, the local environment is close enough to hexagonal so that several quasi-doublet pairs could be expected to be found.

Although attempts have been made to compute the CEF levels in the past for \sdo\ and \sho~\cite{Fennell_2014} as well as more recently for \stmo~\cite{Kademane_2022}, the large number of parameters (caused by the presence of the two inequivalent sites) and the relatively small number of observed \cef\ levels make such computations prone to significant uncertainties.

Despite these difficulties, using a constrained fitting procedure we were able to find a set of \cef\ parameters consistent with the INS spectra and which yielded the observed qualitative magnetisation and susceptibility behaviour.
We should note that these are only one of several nearly equivalent \cef\ parameter sets, albeit the one with the lowest $\chi^2$ statistic.

In all cases we found that we cannot fit three low energy excitations.
In order to fit the intermediate (at 7-13~meV) and high energy ($\approx$30~meV) excitations, and to maintain qualitative agreement with the measured magnetic susceptibility and magnetisation, we found that only schemes with one low energy ($<5$~meV) excitation per Tb site satisfied the data.
As such, we assigned the 0.71 and 1.42~meV levels as ``true'' \cef\ levels whilst the 2.22~meV excitation is deemed to be a dispersive exciton which originates from the same site together with the 0.71~meV mode, as indicated by the lack of intensity at low $Q$ around 0.71~meV.

As each Tb1 (Tb2) ion is coupled to another Tb1 (Tb2) ion in the same unit cell by two different exchange interactions, it is possible to split the same singlet single-ion level into two modes, even in the absence of magnetic ordering, as was observed for the hexagonal U2 sites in UPd$_3$~\cite{Le_2012}.
The lack of dispersion of the 1.42~meV mode thus indicates that the exchange interaction between ions on one of sites is stronger than between ions on the other site.
Despite this hint, however, in order to properly fit the data we should also include the inter-site exchange coupling, which yields a minimum model of five exchange interactions (2 intra-site couplings for each site and the inter-site coupling).
Our powder neutron data, unfortunately are not sufficient to fit such a model, which is the proper subject of future work requiring either single crystal INS measurements of the low energy modes, or a detailed \textit{ab initio} calculation of the exchange interactions.
Thus in the following we fit only the single-ion crystal field parameters. However, we found that after fitting the CEF parameters we were able to use a phenomenological model of the exchange to account for the 2.2~meV peak, although it does not fully match the data.

\begin{table}[tb]
\caption{Crystal field parameters $B^q_p = A^q_p \langle r^p \rangle \theta_p$ (in meV) for the Tb$^{3+}$ ions at two crystallographic sites, Tb1 and Tb2, as determined from fitting the inelastic neutron scattering data.}
\begin{ruledtabular}
\begin{tabular}{lcc}
			&Tb1			&Tb2			\\ \hline 
$B^2_0$		& -0.17741		&  0.055311		\\
$B^2_2$		& -0.094645		&  0.75653		\\
$B^2_{-2}$	&  2.9595e-6	& -5.9839e-5	\\
$B^4_0$		& -0.0074249	&  0.020207		\\
$B^4_2$		& -0.0025211	&  0.0074979	\\
$B^4_{-2}$	& -1.372e-05	&  4.9393e-6	\\
$B^4_4$		&  1.5492e-05	&  4.0977e-6	\\
$B^4_{-4}$	& -0.00011748	&  2.6124e-5	\\ 
$B^6_0$		& -7.3581e-5	&  1.5577e-5	\\
$B^6_2$		&  0.2401		&  0.24772		\\
$B^6_{-2}$	&  0.013588		& -0.017102		\\
$B^6_4$		& -0.011204		& -0.012231		\\
$B^6_{-4}$	& -1.1458e-5	&  3.3862e-5	\\
$B^6_6$		&  0.00013938	& -7.4905e-5	\\
$B^6_{-6}$	&  6.7939e-5	&  3.6555e-5
\end{tabular}
\end{ruledtabular}
\label{cef_pars}
\end{table}  

In order to constrain the fit, each iteration was divided into two steps: 1) the input \cef\ parameters is first refined with respect to the measured energy levels of 0.71, 7.64, and 27.3~meV (Tb1 site), and 1.42, 12.7, and 31.1~meV (Tb2 site) with the wavefunctions (eigenvectors) as variational parameters in an algorithm first proposed by Newman and Ng~\cite{newman_ng_2000}.
The assignment of the energy levels between the two sites was done on the basis of initial parameters from the literature~\cite{Malkin_2015, Malkin_2022} and trial and error.
Energy levels not determined from experimental data were left unchanged in the variational procedure.
2) The peak positions and intensities were calculated from the refined \cef\ parameters in step 1, and a fit of the peak widths was completed yielding a summed $\chi^2$ statistic for the three cuts to the data (the two shown in Fig.~\ref{fig:3_MERLIN} and that in Fig.~S2 in the Supplemental Material~\cite{Supp_2025}). 
Finally a penalty was added to the $\chi^2$ value based on the differences between the measured susceptibility values at 300~K and magnetisation at 80~kOe and those calculated using the \cef\ parameters.
This combined INS $\chi^2$ and physical properties penalty function was then minimised using both local and global search algorithms.

\begin{table*}
\caption{Energy levels from the crystal field parameters in Table~\ref{cef_pars} in meV}
\begin{ruledtabular}
\begin{tabular}{lrrrrrrrrrrrrr}
0 & 0.74 & 7.63 & 8.95 & 17.19 & 28.39 & 29.32 & 43.45 & 45.18 & 50.64 & 50.87 & 54.90 & 56.96 \\
0 & 1.43 & 12.46 & 15.23 & 27.29 & 30.22 & 45.95 & 63.24 & 63.66 & 80.19 & 84.84 & 89.02 & 91.98 \\
\end{tabular}
\end{ruledtabular}
\label{cef_energies}
\end{table*}

The two step iteration ensured that only the \cef\ parameters themselves are varied between iterations. The first step also constrains the energy levels to be close to that measured, while the second step means that the peak widths (which are not resolution limited due to the exchange interaction) should not affect the fitting of the \cef\ parameters.

Starting from the \cef\ parameters in the literature~\cite{Malkin_2015, Malkin_2022, Malkin_2024}, we then refined the $B^2_0$ and $B^4_0$ parameters on each site by trial and error to obtain energy level and intensity schemes which resemble the data.
We then perform a local minimization using Powell's method as implemented in the Scipy Optimize package~\cite{scipy_2020} to find the parameters in Table~\ref{cef_pars}.
In addition, we also explored using a global search with the ``differential evolution'' algorithm as implemented in Scipy to find other starting parameters which were subsequently fitted using  Powell's method but did not find a significantly better set of \cef\ parameters.
We used the Mantid~\cite{mantid68, Arnold_2014} program for the crystal field calculations and the full source code is available for download~\cite{fitscipy}.

The above procedure, and indeed the INS data themselves, can only yield two sets of \cef\ parameters corresponding to the different Tb sites, but cannot explicitly say that a particular set corresponds to a particular site.
We have made the assignment in Table~\ref{cef_pars} based on fitting a point charge model with the Tb1 and Tb2 sites in the positions noted in Table~\ref{tab:W1}.
This assignment yielded the lowest mean absolute error (MAE = $\frac{1}{n}\sum_{p,q} |V^p_q(\mathrm{pc}) - V^p_q(\mathrm{INS})|$ where $n$ is the number of parameters) of 0.32~meV between the normalised point charge model calculated parameters and the normalised INS fitted parameters.
$V^p_q$ denote the ``normalised'' crystal field parameters defined by Fabi~\cite{fabi_focus}.
However, we note that the reverse assignment yielded only a slightly worst MAE of 0.34~meV but required large negative effective point charges for the metallic ions which we believe is physically implausible despite the limitations of the point charge model.
More plausible point charge models yielded larger MAE around 0.45~meV.

With the parameter assignments in Table~\ref{cef_pars}, we found that the calculated magnetic moment for an 80~kOe field along the $c$ direction is 7.3~$\mu_{\rm B}$/Tb for Tb1 and 1.6~$\mu_{\rm B}$/Tb for Tb2 which is in reasonable agreement with the diffraction results in Section~\ref{sec:H_WISH}.

Finally, we mentioned above that we assigned 2.2~meV peak to a dispersive exciton. Using the fitted CEF parameters, and a phenomenological model of the (Heisenberg) exchange, $J(r)=A\cos(2kr)/(2kr)^3$ where $r$ is the distance between pairs of ions and $A=0.4$~meV, $k=0.5$\AA$^{-1}$, we found that a mean-field random-phase-approximation calculation using the McPhase~\cite{mcphase} program could explain the 2.2~meV peak and fit the cut, as shown in Fig.~\ref{fig:3_MERLIN}. We also attempted to use a simpler model, $J(r)=A/r^n$ but this could not fit the data indicating that there is some complexity between the exchange interactions which likely requires single crystal data to unravel.

A more robust approach to the determination of the \cef\ parameters is through a combination of the INS data with the site-selective optical and EPR measurements performed on samples of a nonmagnetic analog, such as SrY$_2$O$_4$ for example, with a small amount of substitution of the magnetic rare earth ion.
Such an approach was used for \seo~\cite{Malkin_2015} and SrY$_2$O$_4$:Er$^{3+}$, where the sets of parameters was derived which fitted all the available experimental data well, including the magnetisation and susceptibility data for dilute and concentrated samples.

According to the latest EPR results~\cite{Malkin_2022}, for Tb1 sites the non-zero component of the g-factor is along the $c$~axis, and the energy levels splitting is similar to what was previously obtained for Er1, Dy1, and Ho1 sites.
The situation is however less clear with the Tb2 sites (the-easy axis direction is near the $b$~axis for them), as the observed optical spectra are inconsistent with the sets of crystal-field parameters used for other members of the family~\cite{Malkin_2022}.

The crystal field parameters obtained from fitting the \sto\ INS data is generally in agreement with the optical spectroscopy and EPR results obtained for dilute SrY$_2$O$_4$:Dy$^{3+}$, SrY$_2$O$_4$:Ho$^{3+}$ and for \seo\ (see Table~I in the Supplemental Material~\cite{Supp_2025}).
However, there are some notable differences, particularly for the $B^6_{\pm 4}$ parameters which are an order of magnitude larger in the INS fit compared to those obtained from optical spectra.
These parameters, though, are crucial to obtain the very low energy levels at 0.71 and 1.42~meV.
With lower values of $B^6_{\pm 4}$, the splitting between the ground state and the first excited state becomes significantly larger ($\approx$2~meV for an order of magnitude reduction in $B^6_{\pm 4}$). 

\subsection{Magnetisation and heat capacity measurements}  \label{sec:mag}
\begin{figure}[tb]
\includegraphics[width=0.99\columnwidth]{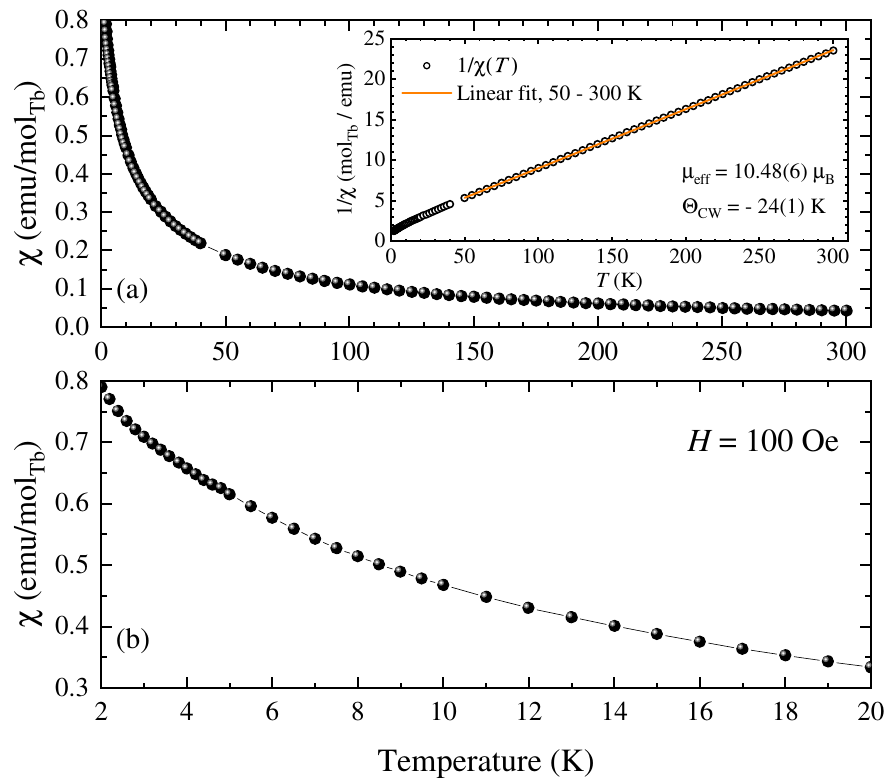}
\caption{Temperature dependence of the magnetic susceptibility, $\chi(T)$, measured on a powder sample of \sto\ in an applied field of 100~Oe.
Panel (a) shows the entire $T$-range covered, 2 to 300~K, while panel (b) focuses on the low-$T$ behaviour emphasising the absence of any discernible features in the range 2 to 20~K.
The inset shows a Curie-Weiss fit to the inverse susceptibility.}
\label{fig:4_chi}
\end{figure} 
Figure~\ref{fig:4_chi} shows the temperature dependence of the magnetic susceptibility $\chi(T)$ of a polycrystalline sample of \sto.
At higher temperatures, the susceptibility follows the Curie-Weiss law, a linear fit in the range 50 to 300~K reveals  $\Theta_{\rm CW}=-24(1)$~K and slightly larger than expected value of the effective magnetic moment of $\mu_{\rm eff}=10.48(6)$~$\mu_{\rm B}$ per \tb\ ion.
This could potentially be explained by the presence of a small frozen field in the superconducting magnet of the MPMS magnetometer (just 10~Oe of frozen field would be sufficient to bring the value of $\mu_{\rm eff}$ below 10~$\mu_{\rm B}$).

A previous publication~\cite{Wu_2020} reported a significant deviation from the Curie-Weiss behaviour below about 160~K and used a temperature range of 180 to 280~K to obtain $\Theta_{\rm CW}=5.00(4)$~K.
The positive (ferromagnetic) sign of $\Theta_{\rm CW}$ in \sto\ would be very surprising given the fact that all seven \sro\ family members for which $\chi(T)$ is known~\cite{Karunadasa_2005,Qureshi_2021_a} have a negative  $\Theta_{\rm CW}$ ranging from about -10 to -100~K.
We note that in Ref.~\cite{Wu_2020}, the susceptibility data were recorded for a ``randomly oriented'' \sto\ crystal, and given a highly anisotropic nature of $\chi(T)$, determination of  $\Theta_{\rm CW}$ from such data is rather unreliable.

Figure~\ref{fig:4_chi}(b) shows the low-temperature (2 to 20~K) behaviour of $\chi(T)$ highlighting a rather smooth variation in the entire region shown, with no evidence for a phase transition around the suggested ordering temperature of 4.28~K~\cite{Li_2014_STO}.  

\begin{figure}[tb]
\includegraphics[width=0.99\columnwidth]{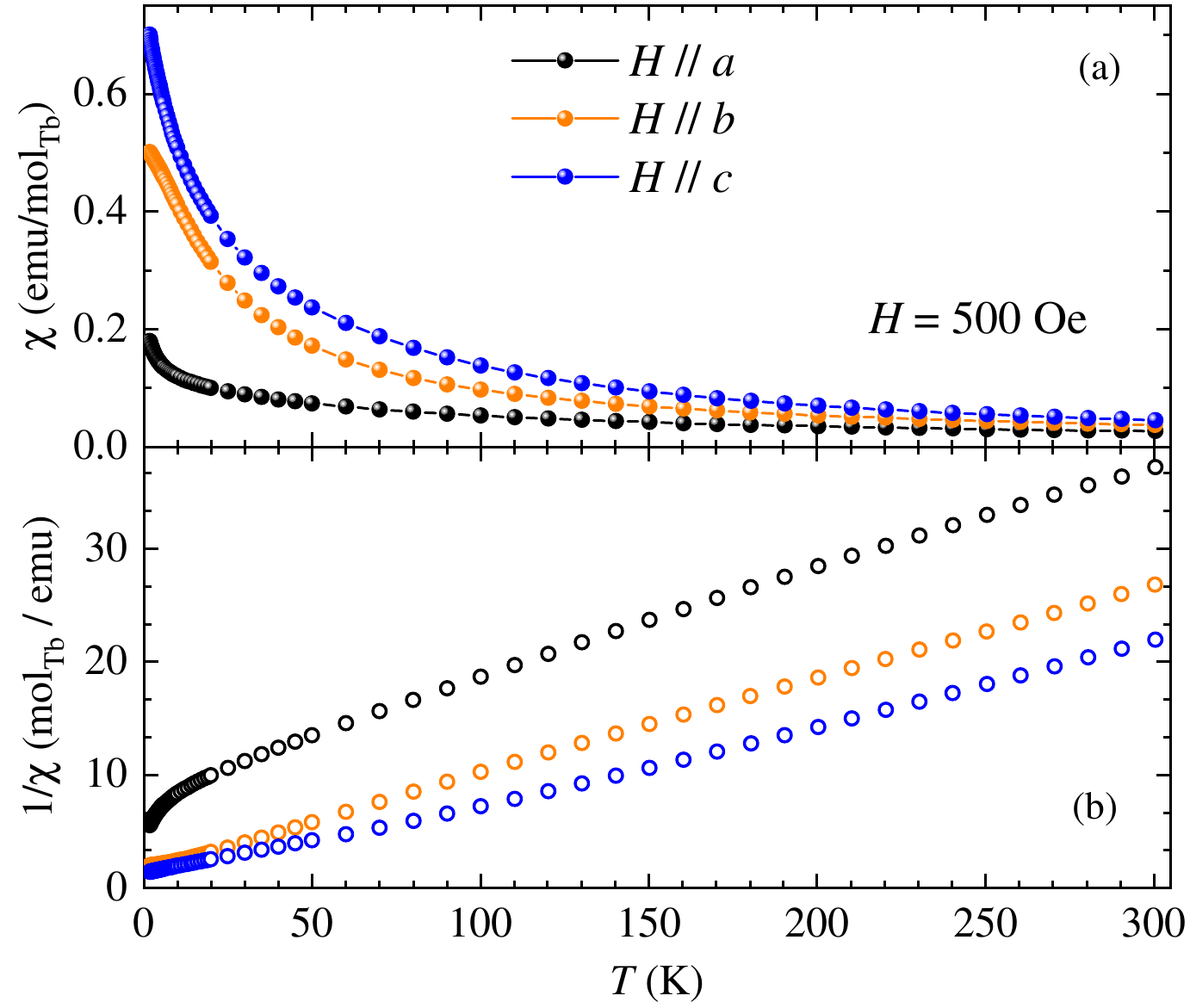}
\caption{Temperature dependence of (a) the magnetic susceptibility, $\chi(T)$ and (b) the inverse susceptibility, $1/\chi(T)$, measured on a single crystal sample of \sto\ in a field of 500~Oe applied along the three principal axes.}
\label{fig:5_chi}
\end{figure} 
The temperature dependence of the magnetic susceptibility measured on a single crystal sample is given in Fig.~\ref{fig:5_chi}a.
The susceptibility is highly anisotropic and is largest for $H \! \parallel \! c$.
The inverse susceptibility [Fig.~\ref{fig:5_chi}(b)] is not linear with temperature for any direction of an applied field, however, when averaged over the three directions, $1/\chi(T)$ does follow the Curie-Weiss law (see figure~S4 in the Supplemental Material~\cite{Supp_2025}).
The non-linearity of the inverse susceptibility as well as the highly anisotropic nature can be explained by the influence of the low-energy CEF levels observed in our INS data.

A linear fit of the averaged inverse susceptibility in the range 50 to 300~K gives a magnetic moment of $\mu_{\rm eff}=9.8(1)$~$\mu_{\rm B}$ per \tb\ ion and $\Theta_{\rm CW}=-44(1)$~K.
Again, the low-temperature behaviour of the $\chi(T)$ curves for all three directions of the applied field is rather smooth giving no reason to suspect a magnetic ordering transition down to at least 2~K.
We note that the recent susceptibility measurements~\cite{DLQC_2019} performed on the sample used in Ref.~\cite{Li_2014_STO} returned results very similar to those shown in Fig.~\ref{fig:4_chi}, that is for the sample for which a magnetic ordering transition was suggested at 4.28~K, the magnetic susceptibility returns no signs of such a transition. 

\begin{figure}[tb]
\includegraphics[width=0.99\columnwidth]{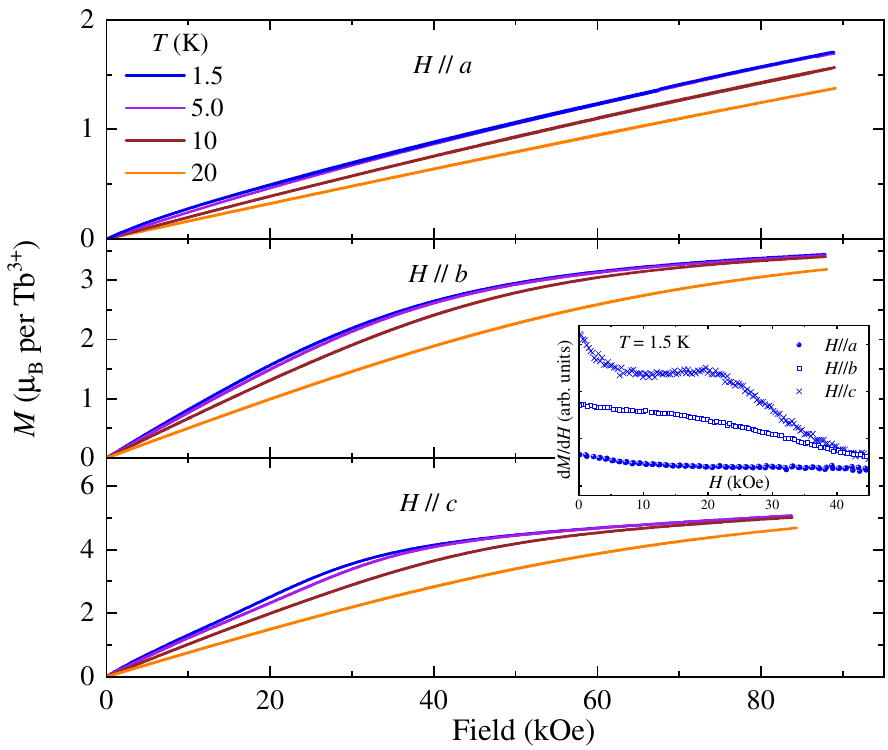}
\caption{Field dependence of magnetisation $M(H)$ of \sto\ single crystal measured at various temperatures for the three principal directions of the magnetic field $H$.
The inset shows field derivatives $dM(H)/dH$ of magnetisation measured at $T=1.5$~K.
The applied field is demagnetisation-corrected.}
\label{fig:6_MH}
\end{figure}

The field dependence of the magnetisation measured on a single crystal sample of \sto\ for $H$ parallel to the three principal crystallographic axes are shown in Fig.~\ref{fig:6_MH}.
For the purposes of calculating the demagnetisation field, the sample was approximated to a rectangular prism with the length along the $c$~axis half that of the other two directions.
The field applied along the $c$~axis returns the highest magnetisation, while the $a$~axis seems to be the hardest to magnetise along.
There is no significant difference between the magnetisation curves measured at 1.5 and 5.0~K for all three fields directions again emphasising the fact that no ordering occurs between these temperatures.
Magnetisation curves do not follow the Brillouin function expected for a simple paramagnet at any of the temperatures measured, but they are in qualitative agreement with the calculations based on the crystal field parameters shown in Tables~\ref{cef_pars} and ~\ref{cef_energies}.

At low temperatures, for $H \! \parallel \! c$, there is a broad maximum in the derivative $dM(H)/dH$ at around 20~kOe (clearly absent for $H \! \parallel \! a$ and $H \! \parallel \! b$), as shown in the inset in Fig.~\ref{fig:6_MH}.
A non-linear field dependence of the derivative of magnetisation $dM(H)/dH$ has also been observed for \seo, \sdo, \sho, and \syo\ single crystals~\cite{Hayes_2012,Quintero_2012,Gauthier_2017c} as well as power samples~\cite{Karunadasa_2005} at temperatures well above the magnetic ordering in these compounds.
Very often, the features in the $dM(H)/dH$ curves are amplified on further cooling, particularly below the magnetic ordering temperatures.
They indicate the boundaries of the fractional magnetisation plateaus (typically of the 1/3 type associated with the up-up-down structures) or the locations of the metamagnetic transitions. 
Given that \sto\ remains in a magnetically disordered state down to the lowest temperature achieved in this study, it is not surprising that no well-defined magnetisation plateaus are observed in this compound, dominated by the single-ion physics rather than by the exchange interactions.

A large influence of the crystal-field effects is also evident from the fact that only about a half of the full magnetic moment expected for the \tb\ ions is recovered in the highest applied field of 90~kOe.
This situation is fully consistent with our CEF calculations, it is also common for many members of the \sro\ family, as the rare-earth ions at the two crystallographic sites often have nearly orthogonal directions for easy-axis magnetisation, which makes full magnetic polarisation nearly impossible for any field direction.

\begin{figure}[tb]
\includegraphics[width=0.99\columnwidth]{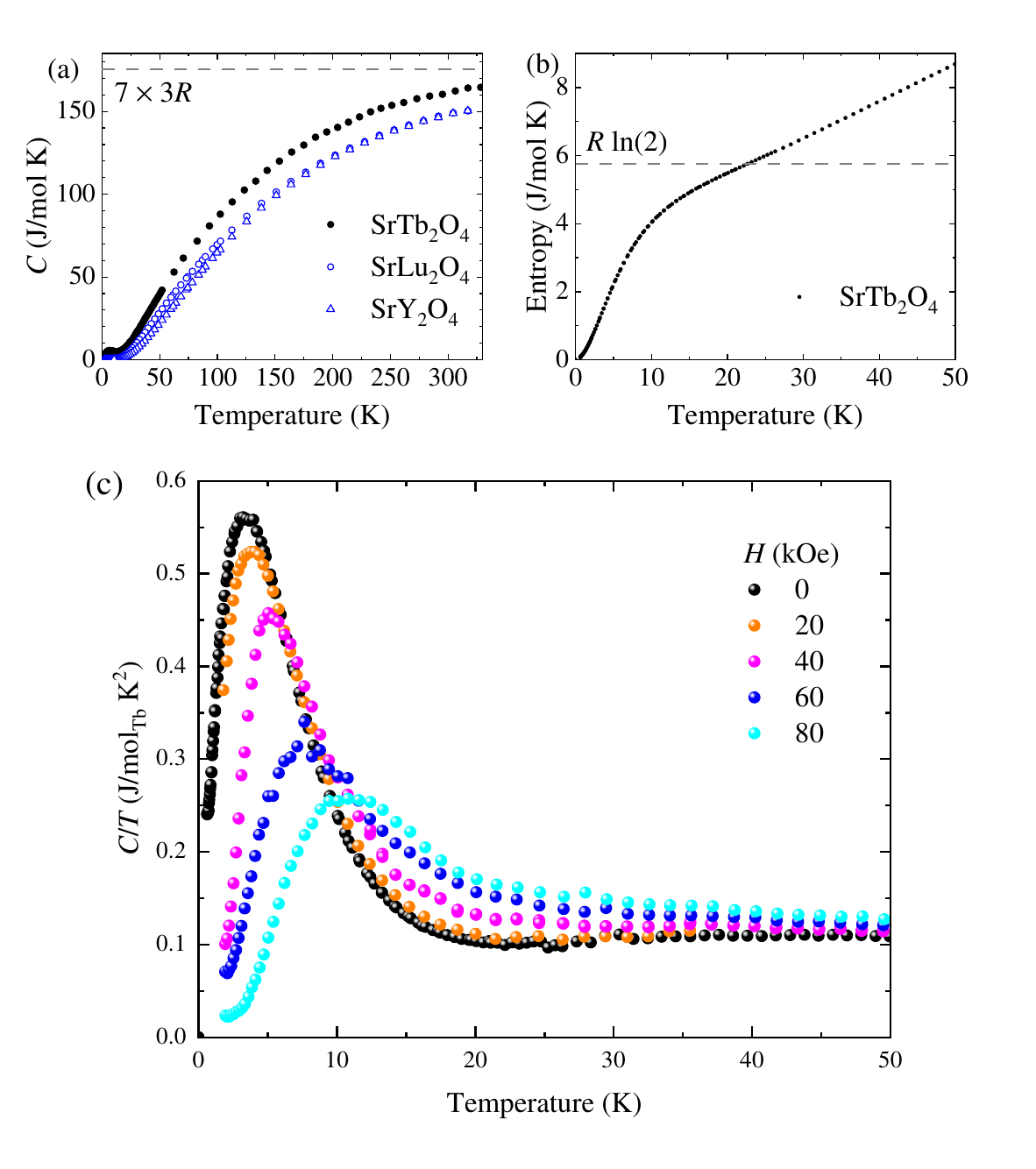}
\caption{(a) Temperature dependence of the heat capacity, $C(T)$, of \sto\ crystal as well as of the isostructural nonmagnetic $\rm SrY_2O_4$ and $\rm SrLu_2O_4$ compounds shown in the entire temperature range measured, 0.6 to 320~K.
(b) Temperature dependence of the magnetic entropy of \sto\ estimated from the $C_{\rm mag}(T)/T$ curve after subtracting a phonon contribution.
(c) $C_{\rm mag}(T)/T$ curves (after the phonon contribution subtraction) measured in different fields applied along the $c$~axis.}
\label{fig:7_HC}
\end{figure} 
Figure~\ref{fig:7_HC} summarises the results of the heat capacity measurements for the \sto\ and the two nonmagnetic isostructural compounds, $\rm SrY_2O_4$ and $\rm SrLu_2O_4$.
For all the compounds, the heat capacity is likely to reach the theoretical maximum, $7 \times 3R$, (7 being the number of atoms in the formula unit) well above room temperature.
From a direct comparison of the heat capacity curves, $C(T)$, [see Fig.~\ref{fig:7_HC}(a)] it is rather obvious that $C(T)$ of \sto\ is systematically higher than the heat capacity of the nonmagnetic blanks in the entire temperature range measured.
The most likely reasons for that are the development of the magnetic correlations, particularly at the lower temperature, as well as the crystal field effects, particularly at the intermediate and higher temperatures. 
Given the relatively small difference in $C(T)$ for $\rm SrY_2O_4$ and $\rm SrLu_2O_4$ and the fact that the Tb ions are much more similar in mass to the Lu ions compared to the Y ones, in all the calculations below we used the $\rm SrLu_2O_4$ data for estimating a phonon contribution to the heat capacity of \sto.

Figure~\ref{fig:7_HC}(b) shows the temperature dependence of the magnetic entropy of \sto\ obtained by integrating the $C_{\rm mag}(T)/T$ curves after the phonon contribution subtraction.
At around $T\approx 22$~K the entropy exceeds the $R\ln (2)$ value expected for a simple model with an effective spin of 1/2 and increases further with increasing temperature.

Figure~\ref{fig:7_HC}(c) demonstrates how the low-temperature part of the $C_{\rm mag}(T)/T$ curves vary on application of the magnetic field along the $c$~axis.
Although the maximum in $C_{\rm mag}(T)/T$ curves remains clearly visible for all fields applied, it shifts to higher temperature with an increasing field and also decreases in amplitude.

None of the curves, including the one measured in zero field, shows an anomaly at 4.28~K, providing further evidence for the absence of a magnetic transition at this temperature or, in fact, at any temperature down to at least 0.6~K (the lowest temperature reached in the heat capacity measurements).

\subsection{Single-crystal neutron diffraction}  \label{sec:scnd}
\subsubsection{Zero field, Laue diffraction}  \label{sec:zeroH_cyclops}

\begin{figure}[tb]
\includegraphics[width=1.25\columnwidth]{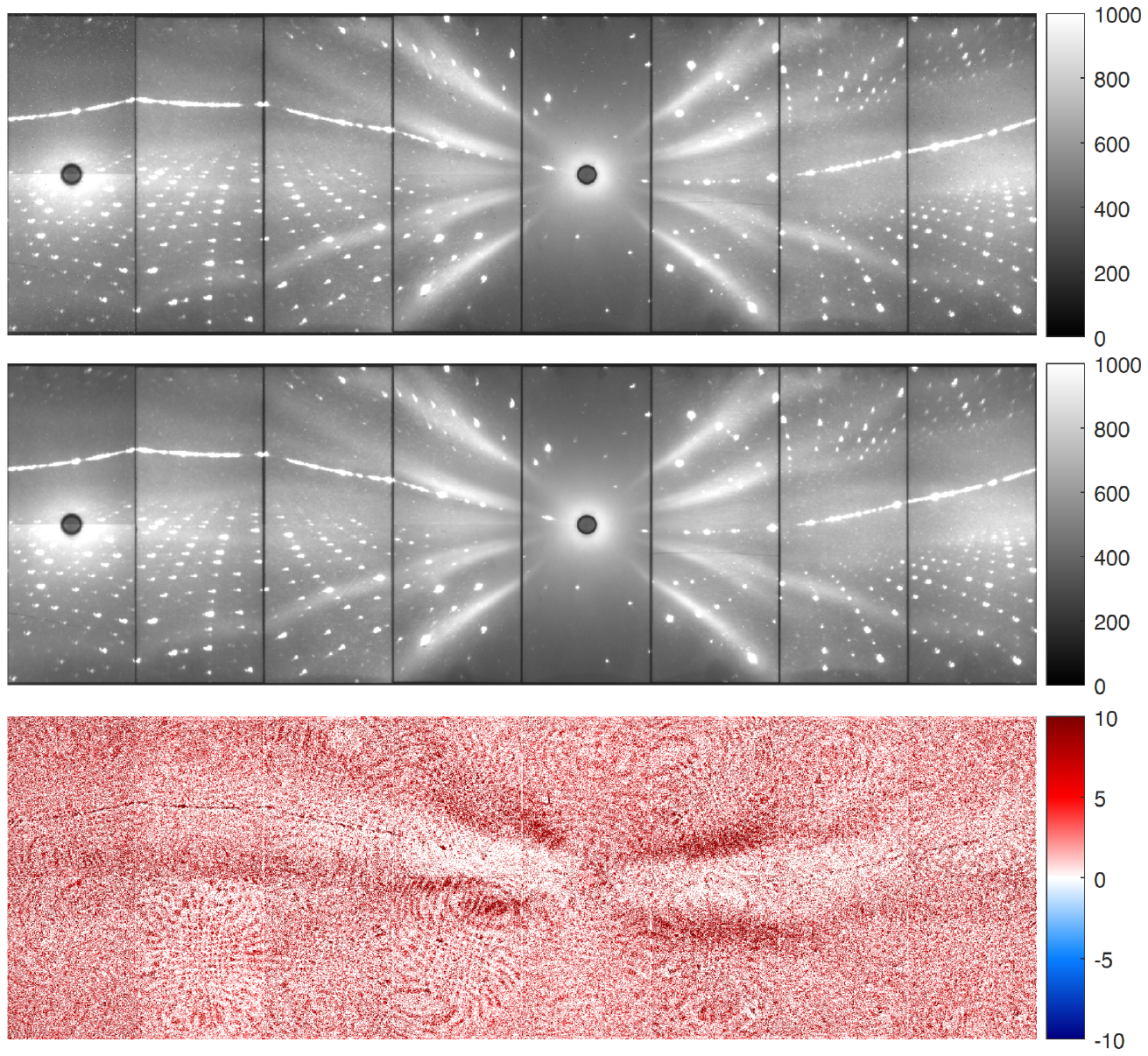}
\caption{Neutron Laue diffraction patterns of \sto\ single crystal taken on the CYCLOPS diffractometer at the ILL at (top) 2.0~K and (middle) 10.0~K.
The bottom panel is a digital difference between the two diffraction patterns shown above.}
\label{fig:8_cyclops}
\end{figure} 
The absence of magnetic ordering in \sto\ is clearly seen from the neutron Laue diffraction patterns depicted in Figure~\ref{fig:8_cyclops}~\cite{CYCLOPS_2018}.
The two patterns taken at 2 and 10~K are identical, as seen in the difference plot (bottom panel of Fig.~\ref{fig:8_cyclops}), and no extra peaks are detected at lower temperature, the intensities of all the Bragg peaks remain constant. 
All the peaks at both 10 and 2~K could be indexed within the nuclear $Pnam$ group.
To make sure that no extra magnetic peaks were missed, we repeated the measurements for a different sample orientation.
The Laue patterns measured at 2 and 10~K (shown in Fig.~S5 in the Supplemental Material~\cite{Supp_2025}) were also identical in this case.
Combining these observations with an exceptionally large coverage of the reciprocal space available on the CYCLOPS diffractometer~\cite{CYCLOPS} one can exclude the presence of any magnetic order with a non-zero propagation vector in \sto\ down to at least $T=2.0$~K.

We note that the powder neutron diffraction pattern reported in Ref.~\cite{Li_2014_STO} contains just a single magnetic Bragg peak at 50~mK, the intensity of this peak is rather weak, about 40\% of the intensity of the nuclear (200) peak, that in turn is an order of magnitude weaker than the main nuclear peaks.
However, even such a weak peak at the incommensurate position should be clearly visible given CYCLOPS' resolution and sensitivity~\cite{CYCLOPS}, since magnetic peaks as weak as 0.3\% of the main nuclear peaks are routinely observed on the diffractometer as long as they have nonzero propagation vector.

\subsubsection{Zero field, polarised neutron diffraction}  \label{sec:zeroH_D7}
\begin{figure}[tb]
\includegraphics[width=1.09\columnwidth]{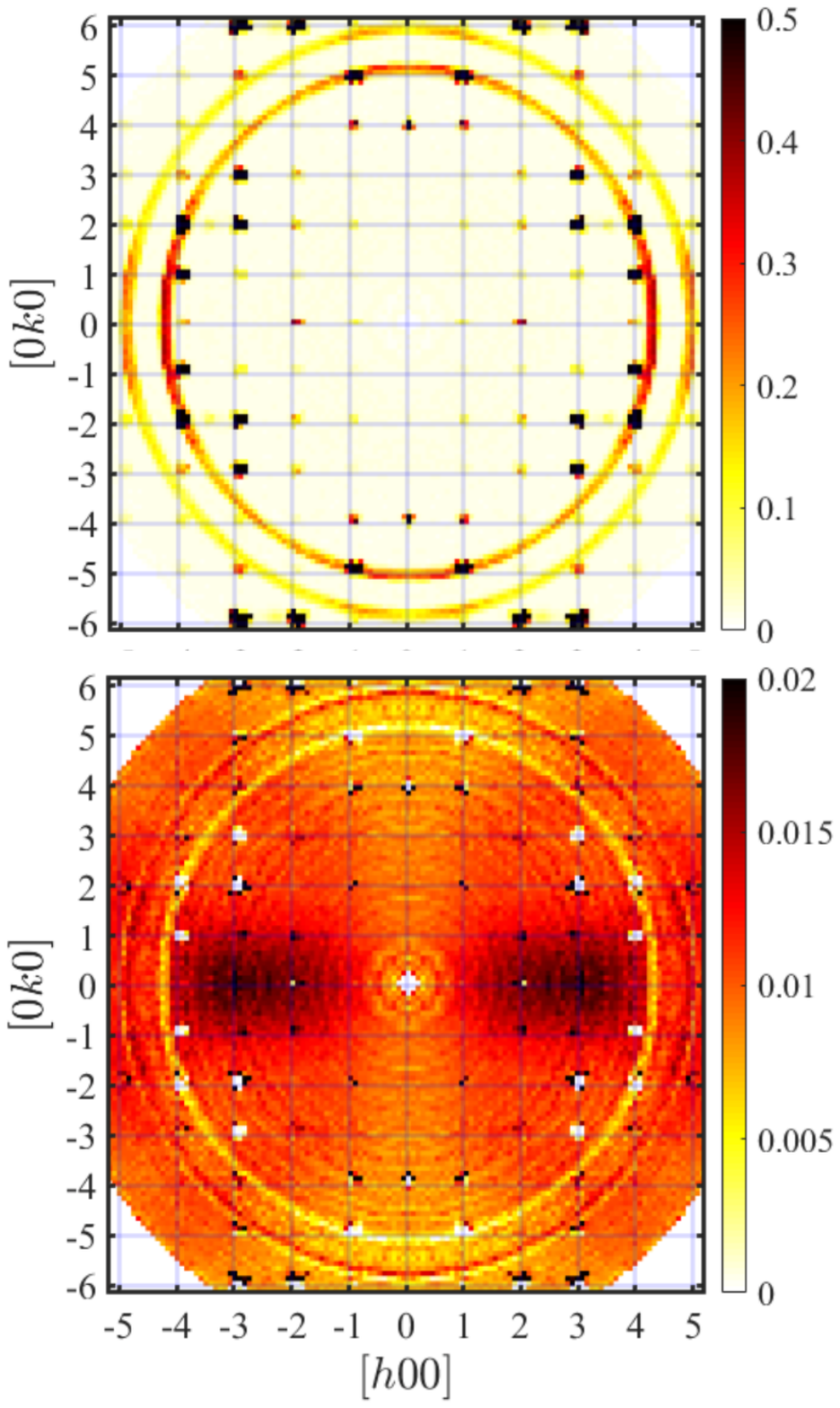}
\caption{Single crystal neutron diffraction maps of the $(hk0)$ plane for \sto\ measured on D7 at $T = 1.5$~K.
The non-spin-flip (top panel) and spin-flip (bottom panel) channels are shown.
To maximise the signal to noise ratio and average over statistics, the intensity map was symmetrized and subsequently unfolded.}
\vspace{-5 mm}
\label{fig:9_D7}
\end{figure}
A central claim of Ref.~\cite{Li_2014_STO} regarding the magnetic ordering in \sto\ is based on the observations made using the diffuse scattering spectrometer D7 with polarisation analysis.
Bragg peaks at ($\pm$1.6~$\pm$1~0) were reported in both SF and NSF channels with the polarization normal to the scattering plane ~\cite{Li_2014_note}.
We have also used the D7 instrument to test for magnetic correlations in a \sto\ crystal.
Figure~\ref{fig:9_D7} shows the intensity maps of the $(hk0)$ scattering plane for the SF and NSF channels at $T=1.5$~K.  The maps were recorded using a polarisation direction along the vertical $c$~axis, normal to the horizontal scattering plane.

The data show some powder rings due to the aluminium sample mount being imperfectly masked with cadmium, however the diffuse background was dominated by forward scattering from windows and ambient neutrons which were correctly estimated from measuring the empty cryostat.
The data in Fig.~\ref{fig:9_D7} have had this diffuse background estimate subtracted, and have been symmetrised by combining data at equivalent $\{hkl\}$ before unfolding.
Unsubtracted and pre-symmetrised data are shown in the Supplemental Material~\cite{Supp_2025}, Fig. S6 and S7.

Our measurements show no evidence for the incommensurate Bragg peaks reported at ($\pm$1.6~$\pm$1~0) in Ref.~\cite{Li_2014_STO}. 
All the Bragg peaks observed in the NSF channel have integer $h$ and $k$.  They are due to nuclear scattering, and their intensities do not change much with temperature on warming up to 50~K (the results of the measurements at selected higher temperatures are given in Fig.~S7 in the Supplemental Material~\cite{Supp_2025}).
Note that for the NSF map, the intensity scale (0 to 0.5~units) is chosen to emphasise the presence of any weaker peaks.  The maximum intensities of the Bragg peaks shown on this map reaches about 5 units.

The intensity scale for the SF map is significantly lower, reflecting the fact that the diffuse scattering intensity is orders of magnitude lower compared to the Bragg peaks.
The SF map also shows very small and localised features at the integer $(hk0)$ positions, some of which are negative.
These are most probably due to a known issue in correcting for polarization inefficiency, stemming from using parameters determined from the diffuse scattering from amorphous quartz to correct sharp, intense, self-collimated Bragg scattering.
These sharp features therefore cannot reliably be attributed to magnetic Bragg peaks from the sample.  

The SF map in Fig.~\ref{fig:9_D7} shows substantial diffuse scattering which is clearly magnetic.
The scattering is most intense around the $\langle 100 \rangle$ directions, particularly for $1.5 < h < 3.5$ and $|k|<1$.
Along the orthogonal $\langle 010 \rangle$ direction, the magnetic diffuse signal is clearly the weakest, while there is a visible intensity increase along the diagonal $\langle 110 \rangle$ directions.
 
The observed diffuse scattering in Fig.~\ref{fig:9_D7} is very different to that observed in Ref.~\cite{Li_2014_STO}, and also to those observed in 
 in the sister compounds \seo\ and \sho~\cite{Hayes_2011,Ghosh_2011,Young_2013,Wen_2015,Young_2019}.
In \seo, the magnetic moments on the Er1 site form a ${\bf q}=0$ structure clearly marked by the sharp magnetic peaks appearing below 0.75~K.
At low temperatures, only a rather weak diffuse scattering is seen for this compound around the magnetic Bragg peaks, the intensity of the diffuse signal rapidly increases on warming above 0.75~K and forms a characteristic lozenge pattern~\cite{Hayes_2011}.  
In \sho, because of the short-range nature of the ordering, the magnetic ${\bf q}=0$ peaks remain broad even at the lowest temperature again forming the same pattern~\cite{Young_2013}.
Figure~\ref{fig:9_D7} shows that such well-developed patterns of diffuse scattering are absent in our sample of \sto, the diffuse scattering which is much broader and not particularly structured.
The diffuse scattering depicted in Fig.~1 of Ref.~\cite{Li_2014_STO} for their sample of \sto\ is also not centred around the ${\bf q}=0$ positions and lacks the characteristic structure seen in \seo\ and \sho.

The incident neutrons of $\lambda =3.1$~\AA\ used on D7 have an energy of 8.51~meV.
As discussed in Section~\ref{sec:INS}, the lower-energy crystal field excitations in \sto\ range from 0.71 to 2.22~meV [see Fig.~\ref{fig:3_MERLIN}(b)].
For these excitations, the quasi-static approximation works rather well. 
There is also a crystal field level at 7.64~meV [see Fig.~\ref{fig:3_MERLIN}(a)] for which the integration over the spin-fluctuation spectrum will be incomplete and it is possible that the entire paramagnetic signal was not observed within the given energy window.
However, for the single crystal measurements, a breakdown of the quasi-static approximation is typically seen as characteristic deformations of the diffuse scattering, resulting in a loss of symmetry when compared to the underlying lattice.
As such deformations are not seen in the scattering maps depicted in Fig.~\ref{fig:9_D7}, we therefore rule out a significant influence of the crystal field effects on the magnetic diffuse scattering observed in \sto.  

\subsubsection{Neutron diffraction in applied field}  \label{sec:H_WISH}
\begin{figure}[tb]
\includegraphics[width=1.15\columnwidth]{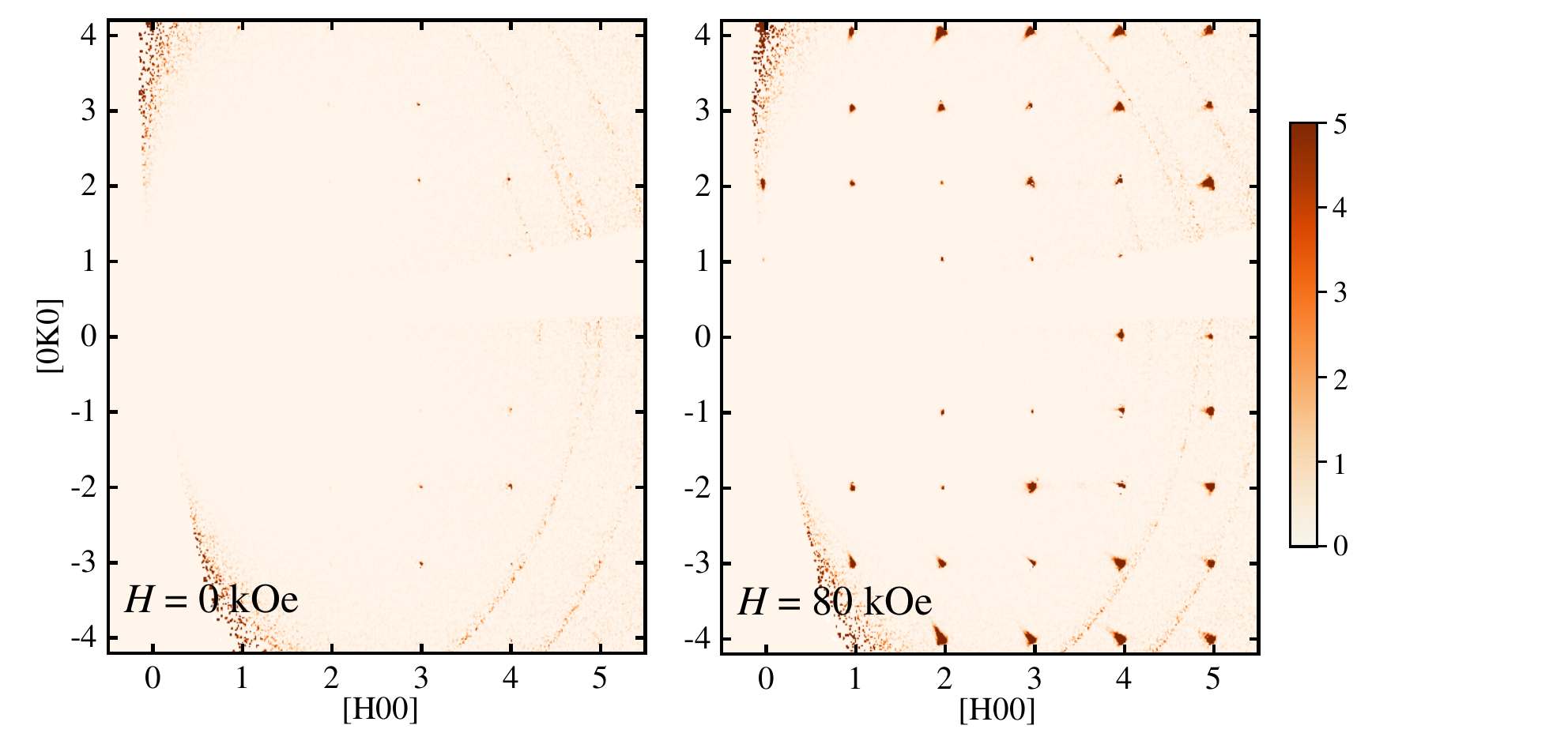}
\caption{Single crystal neutron diffraction maps of the $(hk0)$ plane for \sto\ measured on WISH at $T=35$~mK (left) in zero field and (right) in a field of 80~kOe, $H \! \parallel \! c$.
The magnetic signal is isolated by subtracting a 10~K background.}
\label{fig:10_WISH}
\end{figure} 
The single crystal neutron diffraction experiment on the WISH instrument was conducted with two objectives in mind.

\begin{figure}[tb]
\includegraphics[width=0.92\columnwidth]{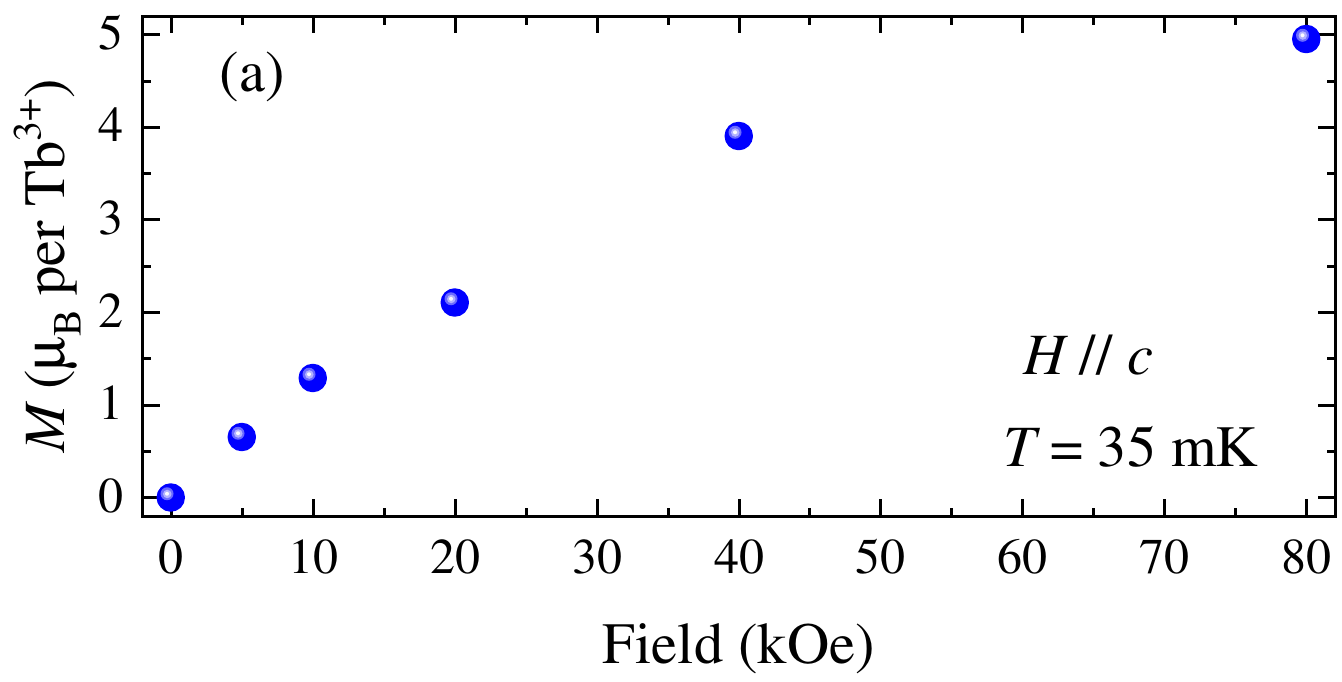}
\includegraphics[width=0.98\columnwidth]{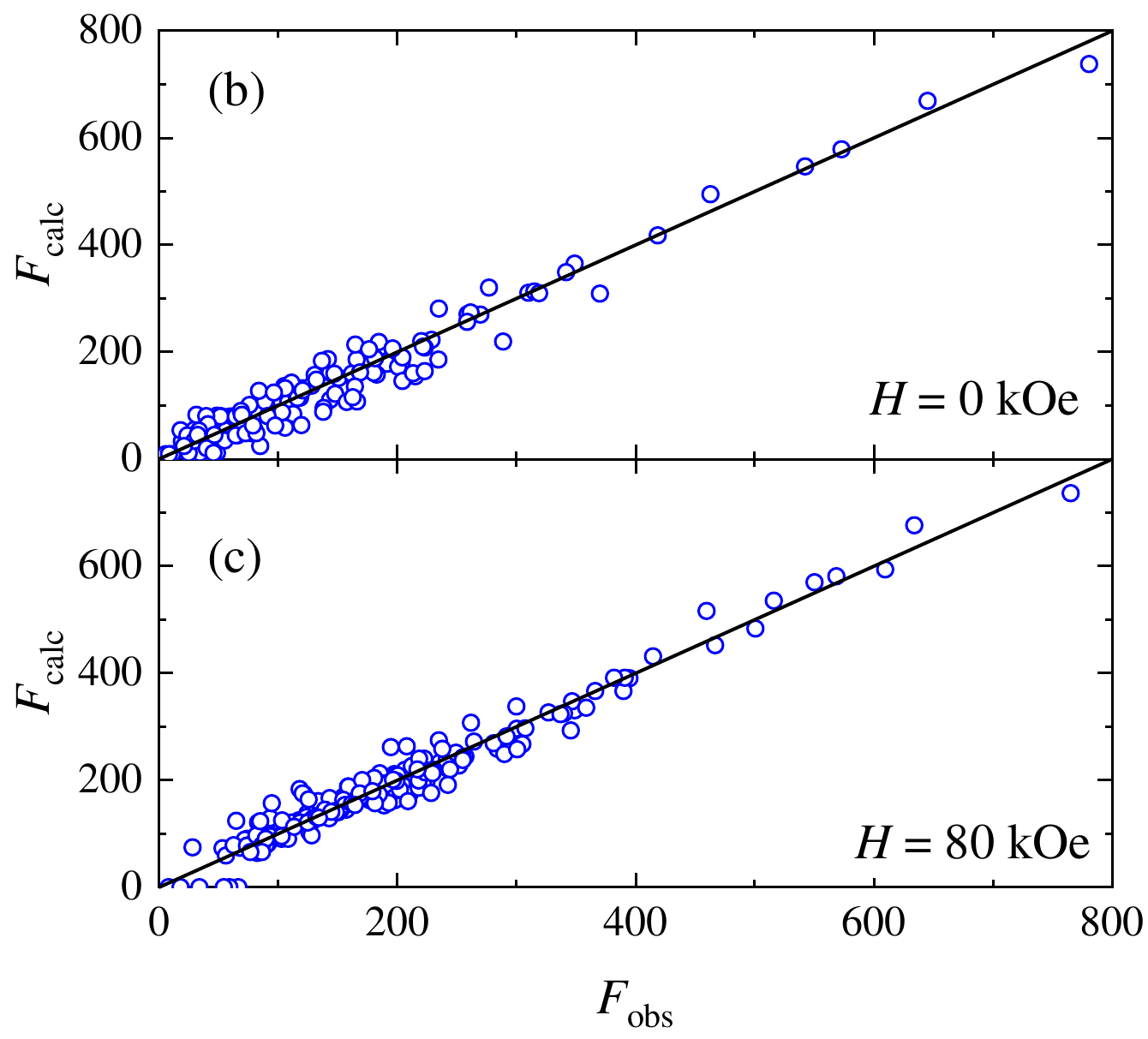}
\caption{(a) Evolution of the total magnetic moment per \tb\ ion as observed by neutron diffraction at $T=35$~mK. 
		Observed structure factors versus calculated one for the single crystal WISH data collected  in (b) 0~kOe and (c) 80~kOe, the error-bars are smaller than the marker size.}
\label{fig:11_WISH}
\end{figure}

Firstly, we wanted to extend the investigations down to dilution cryostat temperatures.
On cooling down to $T=35$~mK, the lowest experimentally available temperature, the intensity of all peaks remained practically unchanged and no new reflections were found in the $(hk0)$ scattering plane, see Fig.~\ref{fig:10_WISH}, ruling out a magnetic ordering transition in zero field.
This observation is further reinforced by the refinement of the extracted single crystal intensities.
Figure~\ref{fig:11_WISH}(b) shows the observed structure factors against the calculated ones for the nuclear structure refinement using the $Pnam.1'$ model for \sto.
The agreement is quite satisfactory, also compared with the x-ray refinements reported in Table~\ref{tab:Xrays}, and the crystal parameters obtained and the reliability factors are reported in Table~\ref{tab:W1}.
Due to the limited number of reflections and the geometry constraints of the cryomagnet which allowed the collection in only the $(hk0)$ plane, we refined isotropic atomic displacement parameters (ADPs) for all the atoms and constrained ADPs for all oxygen sites to the same value.

\begin{table}
\caption{Crystal data, reliability factors and structural parameters for the single crystal refinements of the neutron diffraction data collected on WISH at 0 and 80~kOe.
The atoms sit at the 4c Wyckoff position (x,y,1/4).}
\label{tab:W1}
\begin{tabular}{c|c|c}
\hline 
\hline  
Magnetic Field				& 0~kOe				& 80~kOe		\\
Space group				    & $Pnam.1'$			& $Pn'a'm$	    \\
$a$, $b$, $c$ (\AA)				& \multicolumn{2}{c}{10.0(1), 12.9(1), 3.45(10)} \\
$R_p$					    & 0.155				& 0.0899		\\
$R_{wp}$					& 0.147				& 0.1043		\\
\# reflections				& 154				& 191		    \\
Sr x,  y					& ~0.749(1) 0.646(1)	& ~0.750(1) 0.647(2) \\
Sr $U_{\rm iso}$ (\AA$^2$)	& 0.014(6)   		& 0.008(6)		\\
O1 x, y					    & ~0.216(2) 0.178(2)	& ~0.218(2) 0.175(2)   \\
O2 x, y					    & ~0.130(2) 0.483(1)	& ~0.128(2) 0.482(2)\\
O3 x, y					    & ~0.514(1) 0.781(1)	& ~0.510(2) 0.781(2) \\
O4 x, y					    & ~0.430(1) 0.421(1)	& ~0.427(2) 0.421(2)\\
O1,2,3,4 $U_{\rm iso}$ (\AA$^2$) & 0.020(5)		& 0.014(4)		\\
Tb1 x, y					& ~0.426(1) 0.115(1)	& ~0.425(1) 0.113(1)  \\
Tb1 $U_{\rm iso}$ (\AA$^2$) & ~0.014(5)				& ~0.011(3)		\\
Tb1 $M_{z} (\mu_{\rm B})$   & 0          				& 7.4(3)		\\
Tb2 x, y					& ~0.418(1) 0.612(1)	& ~0.418(1) 0.611(1) \\
Tb2 $U_{\rm iso}$ (\AA$^2$)	& 0.012(5)   		& 0.011(3)		\\
Tb2 $M_{z} (\mu_{\rm B})$   & 0          		& 2.5(2)		\\              
\hline                      
\hline
 \end{tabular}
\end{table}

The second aim for the WISH experiment was to investigate the influence of applied field. 
The magnetic field was applied along the $c$~axis at $T=35$~mK, the measurements were taken in fields of 5, 10, 20, 40, and 80~kOe.
We have observed a significant increase in the intensity of most of the $(hk0)$ peaks with integer $h$ and $k$ in the applied field.
Figure~\ref{fig:10_WISH} presents the background-subtracted intensity maps in zero field and in the maximum field of 80~kOe clearly showing an increase in the intensity of the ${\bf q}=0$ peaks.
Although the high-$T$ background subtraction is not perfect (as for example, the powder lines from the sample holder are still visible after subtraction), it is rather obvious that the field-induced magnetic intensity is orders of magnitude higher than any artefacts.

The data collected in 80~kOe were integrated and refined assuming a ferromagnetic structure with Tb moments aligned along the $c$~direction.
This corresponds to the magnetic space group $Pn'a'm$ derived from the $m\Gamma_4^+$ irreducible representation.
The single crystal refinement returned a good agreement between the observed and calculated structure factors as can be appreciated from Fig.~\ref{fig:11_WISH}(c).
Atomic position, ADPs and magnetic moments for the two Tb ions (constrained along the field direction) were refined and the results are reported in Table~\ref{tab:W1}.
The refinement converged with different magnetic moments values for the two Tb positions: 7.4(3) and 2.5(2)$\mu_{\rm B}$ for Tb1 and Tb2, respectively.
Attempts to constrain the two sites to have the same moment size or different moment ratio returned significantly worse fits.
The predicted intensity maps look very similar for the models with only Tb1 or Tb2 having a magnetic moment on them, but rather different if both sites carry the same moment (see Fig.~S8 in the Supplemental Material~\cite{Supp_2025}). 

The different moment size of the two RE sites is related to their different \cef\ environments, as observed in several other \sro\ compounds~\cite{Petrenko_2008,Qureshi_2021_a,Qureshi_2022}. This is also fully consistent with our calculations based on the crystal field parameters shown in Tables~\ref{cef_pars}, which returned very different sizes for the Tb1 and Tb2 moments in 80~kOe of magnetic field applied along the $c$~axis.

The observed total ferromagnetic moment as function of applied field [see Fig.~\ref{fig:11_WISH}(a)] is consistent with the magnetisation data shown in Fig.~\ref{fig:6_MH} despite having been taken at different temperatures, 35~mK and 1.5~K respectively.
This is not surprising, as for $H \! \parallel \! c$, the magnetisation in 80~kOe reaches saturation and therefore is practically temperature independent for $T<10$~K.
 
\section{Summary}

To summarise, we have shown that \sto\ does not order magnetically down to at least 35~mK.
Various results from neutron diffraction, magnetic susceptibility, and heat capacity measurements definitively rule out a transition at 4.28~K contrary to previous reports.
The nonmagnetic ground state of \sto\ is governed by an intricate balance of the geometrically frustrated exchange interactions between the \tb\ ions and the splitting of the energy levels by the crystal fields acting on them.  

Using a constrained search procedure, we were able to obtain a set of crystal field parameters consistent with previous work on other \sro\ systems and which also qualitatively agrees with the measured susceptibility and magnetisation (predicting that the $c$~axis is the easy axis and the $a$~axis is the hard axis), and which also agrees with the ordered moments of the two Tb sites as determined by neutron diffraction.
However, we should note that due to the large number of symmetry allowed crystal field parameters, the fitting problem is quite unconstrained, and that the parameters we obtained may not necessarily be the correct values.

\section*{ACKNOWLEDGMENTS}
We are grateful to B.~F\aa k and B.Z.~Malkin for discussions, as well as to D.L.~Quintero-Castro for sharing unpublished data.
The authors would like to acknowledge the contributions of D.L.~Elmer and M.J.J.~Johnson to the preparation of \sto\ polycrystalline and crystal samples through their participation in the undergraduate project.
We would also like to acknowledge the expertise and dedication of the low-temperature groups at both the Institut Laue-Langevin and ISIS.
The work at the University of Warwick was supported by EPSRC through grants EP/M028771/1 and EP/T005963/1.
For the purpose of open access, the authors have applied a Creative Commons Attribution (CC-BY) licence to any author accepted manuscript version arising.
All data underlying this work are available from the authors upon reasonable request.
	
\bibliography{SrLn2O4_all}
\end{document}


\renewcommand{\thefigure}{S\arabic{figure}}
\renewcommand{\thetable}{S\arabic{table}}
\renewcommand{\theequation}{S\arabic{equation}}

\makeatletter
\renewcommand\@bibitem[1]{\item\if@filesw \immediate\write\@auxout
    {\string\bibcite{#1}{S\the\value{\@listctr}}}\fi\ignorespaces}
\def\@biblabel#1{[S#1]}
\makeatother

\date{\today}
\title{Supplemental Material to ``Magnetic properties of the zigzag ladder compound SrTb$_2$O$_4$''}
	\author{F. Orlandi}			\affiliation{ISIS Neutron and Muon Source, STFC Rutherford Appleton Laboratory, Chilton, Didcot, OX11 0QX, United Kingdom}
	\author{M.~Ciomaga~Hatnean} \affiliation{Department of Physics, University of Warwick, Coventry, CV4 7AL, United Kingdom}
	\author{D.A.~Mayoh}			\affiliation{Department of Physics, University of Warwick, Coventry, CV4 7AL, United Kingdom}
	\author{J.P.~Tidey}			\affiliation{Department of Chemistry, University of Warwick, Coventry, CV4 7AL, UK}
	\author{S.X.M.~Riberolles}	\affiliation{Department of Physics, University of Warwick, Coventry, CV4 7AL, United Kingdom}
	\author{G.~Balakrishnan}		\affiliation{Department of Physics, University of Warwick, Coventry, CV4 7AL, United Kingdom}											
	\author{P.~Manuel}			\affiliation{ISIS Neutron and Muon Source, STFC Rutherford Appleton Laboratory, Chilton, Didcot, OX11 0QX, United Kingdom}
	\author{D.D.~Khalyavin}		\affiliation{ISIS Neutron and Muon Source, STFC Rutherford Appleton Laboratory, Chilton, Didcot, OX11 0QX, United Kingdom}
	\author{H.C.~Walker}		\affiliation{ISIS Neutron and Muon Source, STFC Rutherford Appleton Laboratory, Chilton, Didcot, OX11 0QX, United Kingdom}
	\author{M.D.~Le}			\affiliation{ISIS Neutron and Muon Source, STFC Rutherford Appleton Laboratory, Chilton, Didcot, OX11 0QX, United Kingdom}													
	\author{B.~Ouladdiaf}		\affiliation{Institut Laue-Langevin, 71 Avenue des Martyrs, CS 20156, 38042 Grenoble Cedex 9, France}
	\author{A. R. Wildes}			\affiliation{Institut Laue-Langevin, 71 Avenue des Martyrs, CS 20156, 38042 Grenoble Cedex 9, France}
	\author{N.~Qureshi}			\affiliation{Institut Laue-Langevin, 71 Avenue des Martyrs, CS 20156, 38042 Grenoble Cedex 9, France}			
	\author{O.A.~Petrenko}		\affiliation{Department of Physics, University of Warwick, Coventry, CV4 7AL, United Kingdom}

\maketitle
\subsection{Samples preparation} 
\begin{figure}[tb]
\includegraphics[width=0.99\columnwidth]{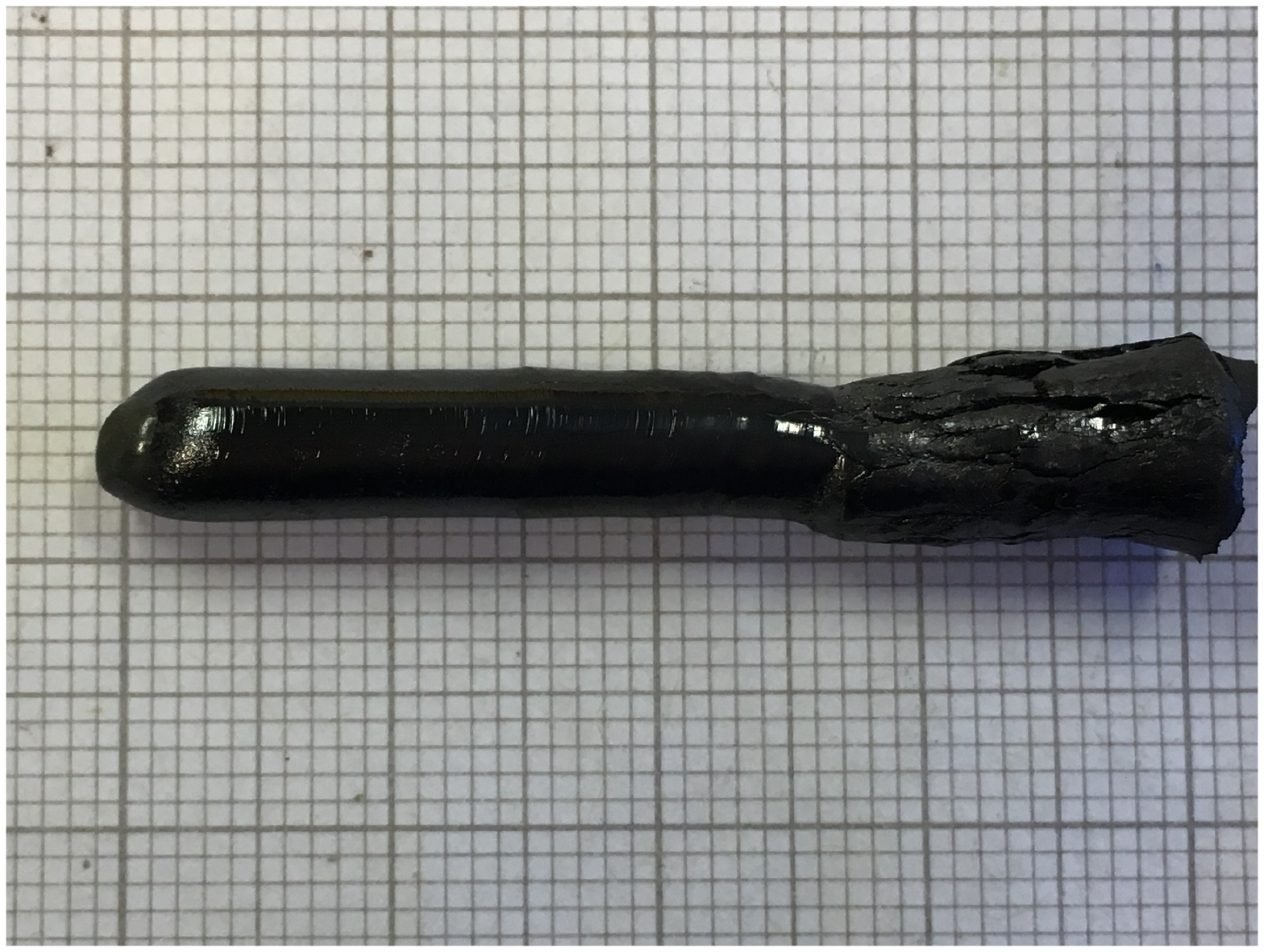}
\caption{Boule of \sto\ prepared by the floating-zone method in an high purity argon atmosphere with a growth rate of 4~mm/h.}
\label{fig:xtal}
\end{figure}

Figure~\ref{fig:xtal} shows a photograph of the as-grown boule of \sto.

\subsection{Inelastic neutron scattering}

\begin{figure}[tb]
\includegraphics[width=0.99\columnwidth]{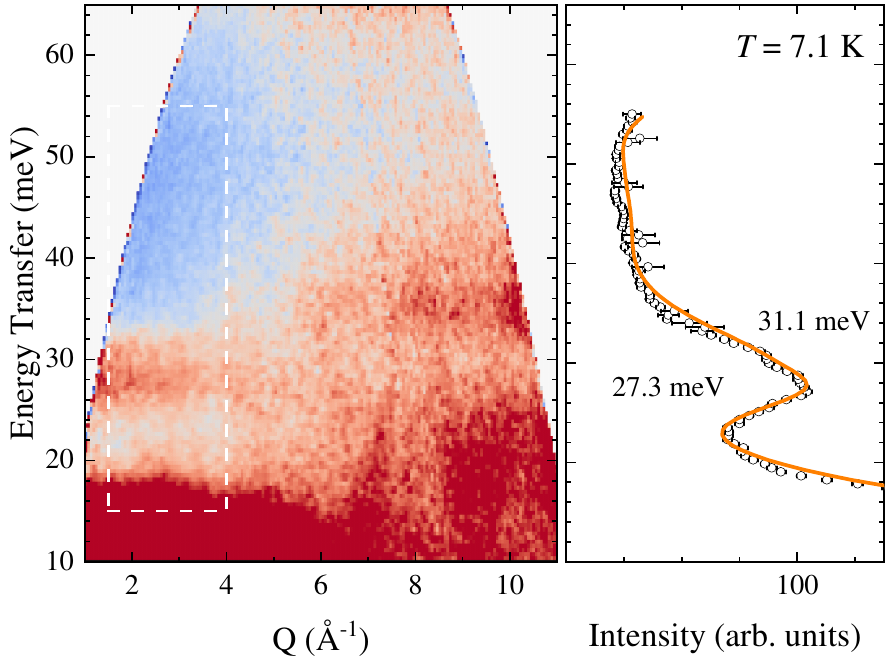}
\caption{INS data for \sto\ powder sample measured on MERLIN at $T = 7.1$~K with incident energy neutrons of $E_i=82$~meV.
To identify the exact energies of the observed excitations, the cuts along the energy transfer are shown to the right of the colour intensity map.
The experimental data are shown by black symbols whilst the best fit to the CEF Hamiltonian as described in the main text is shown as a solid orange line.
White dashed lines on the colour map indicate the area used for the cut.}
\label{fig:MERLIN_82meV}
\end{figure} 

Inelastic neutron scattering spectrum measured on the MERLIN spectrometer at $T=7.1$~K with incident energy neutrons of $E_i=82$~meV is shown in Fig.~\ref{fig:MERLIN_82meV}.
There are two partially overlapping crystal field excitations at 27.3 and 31.1~meV.
The signal at higher scattering vector, $Q>6$~\AA$^{-1}$ is due to the phonons.

\begin{figure*}[tb]
\includegraphics[width=2.0\columnwidth]{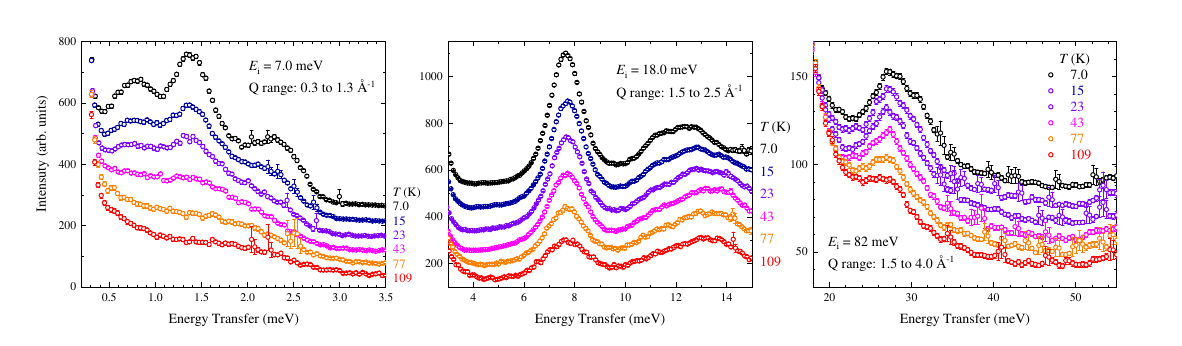}
\caption{Temperature evolution of the $S(Q,\omega)$ for \sto\ powder sample measured on MERLIN with the incident energy neutrons of (left) 7, (centre) 18, and (right) 82~meV.
The curves are consecutively offset for clarity.}
\label{fig:MERLIN_Temperature}
\end{figure*}

Figure~\ref{fig:MERLIN_Temperature} illustrates the temperature dependence of the INS data for three different incident energy neutrons.


\subsection{Crystal electric field analysis}

Table~\ref{TableSI} lists the crystal-field parameters for the rare-earth ions in \sto, \seo, $\rm SrY_2O_4$:Ho$^{3+}$, and $\rm SrY_2O_4$:Dy$^{3+}$.

Note that for ease of comparison between different ions we use a different convention here than that used in the main text. Although we use the ``Stevens normalisation'' in both, here we define the parameters as $A^q_p \langle r^p \rangle$ whereas in the main text we use $B^q_p = A^q_p \langle r^p \rangle \theta_p$ where the Stevens operator equivalent factors $\theta_p$ are included in the parameters. Although the $B^q_p = A^q_p \langle r^p \rangle \theta_p$ convention is the most commonly used in the neutron scattering literature, including the $\theta_p$ factors means that there would be large numerical differences due to the different ground state multiplets between the different ions, making comparison of the ``intrinsic'' parameters between the ions more difficult.


\begin{table*}[tb] 
\caption{Crystal field parameters $A^q_p \langle r^p \rangle$ (in meV) for the Tb$^{3+}$ ions at two crystallographic sites, Tb1 and Tb2 as determined from the inelastic neutron scattering data fitting in this work.
For comparison, we also list the parameters for Er$^{3+}$ in \seo~\cite{Malkin_2015}, Ho$^{3+}$ in $\rm SrY_2O_4$~\cite{Nikitin_2023} and Dy$^{3+}$ in $\rm SrY_2O_4$~\cite{Malkin_2024}.}
\begin{ruledtabular}
\begin{tabular}{lr|cccc|cccc}
\multicolumn{2}{c|}{}	&  \multicolumn{4}{c|}{Position 1}		& \multicolumn{4}{c}{Position 2} 		\\  
$p$	&$q$&Tb1	&Dy1	& Ho1	& Er1	&Tb2	&Dy2	& Ho2	& Er2	\\ \hline 
2	&0	&17.56	&22.51	&24.84	&14.63	&-5.476	&-0.62	&-0.992	&2.108	\\
2	&2	&9.370	&11.18	&17.74	&17.05	&-74.90	&-90.40	&-92.75	&-92.26	\\
2	&-2	&-23.77	&-14.10	&-17.68	&-21.23	&-24.52	&-17.98	&-16.49	&-15.50	\\
4	&0	&0.024	&-7.961	&-7.372	&-7.105	&-0.489	&-8.023	&-7.812	&-7.465	\\
4	&2	&-60.64	&-133.2	&-132.5	&-132.2	&-139.7	&136.8	&136.4	&128.1	\\
4	&-2	&110.98	&146.3	&147.1	&144.5	&165.0	&-115.0	&-121.6	&-121.2	\\
4	&4	&-20.59	&-9.759	&-7.738	&-10.78	&-99.89	&47.14	&50.59	&53.34	\\
4	&-4	&-91.51	&-120.6	&-116.8	&-120.6	&61.24	&-95.55	&-88.66	&-85.01	\\ 
6	&0	&12.24	&-5.171	&-5.078	&-4.712	&-4.405	&-4.613	&-4.576	&-4.365	\\
6	&2	&-13.82	&-4.960	&-2.740	&-2.765	&-3.655	&-8.854	&-8.680	&-8.482	\\
6	&-2	&10.22	&5.369	&2.864	&2.827	&-30.20	&-4.687	&-4.638	&-5.307	\\
6	&4	&104.8	&3.063	&0.4712	&3.732	&-23.30	&-9.945	&-9.052	&-9.945	\\
6	&-4	&-124.3	&-17.60	&-18.74	&-14.28	&66.81	&-26.21	&-25.79	&-23.73	\\
6	&6	&65.63	&-21.12	&-19.31	&-20.11	&-13.89	&-14.83	&-14.26	&-14.83	\\
6	&-6	&-60.59	&-11.07	&-12.31	&-10.42	&-32.60	&11.22	&11.78	&9.982 
\end{tabular}
\end{ruledtabular}
\label{TableSI}
\end{table*}



\subsection{Magnetic susceptibility measurements}

\begin{figure}[tb]
\includegraphics[width=0.95\columnwidth]{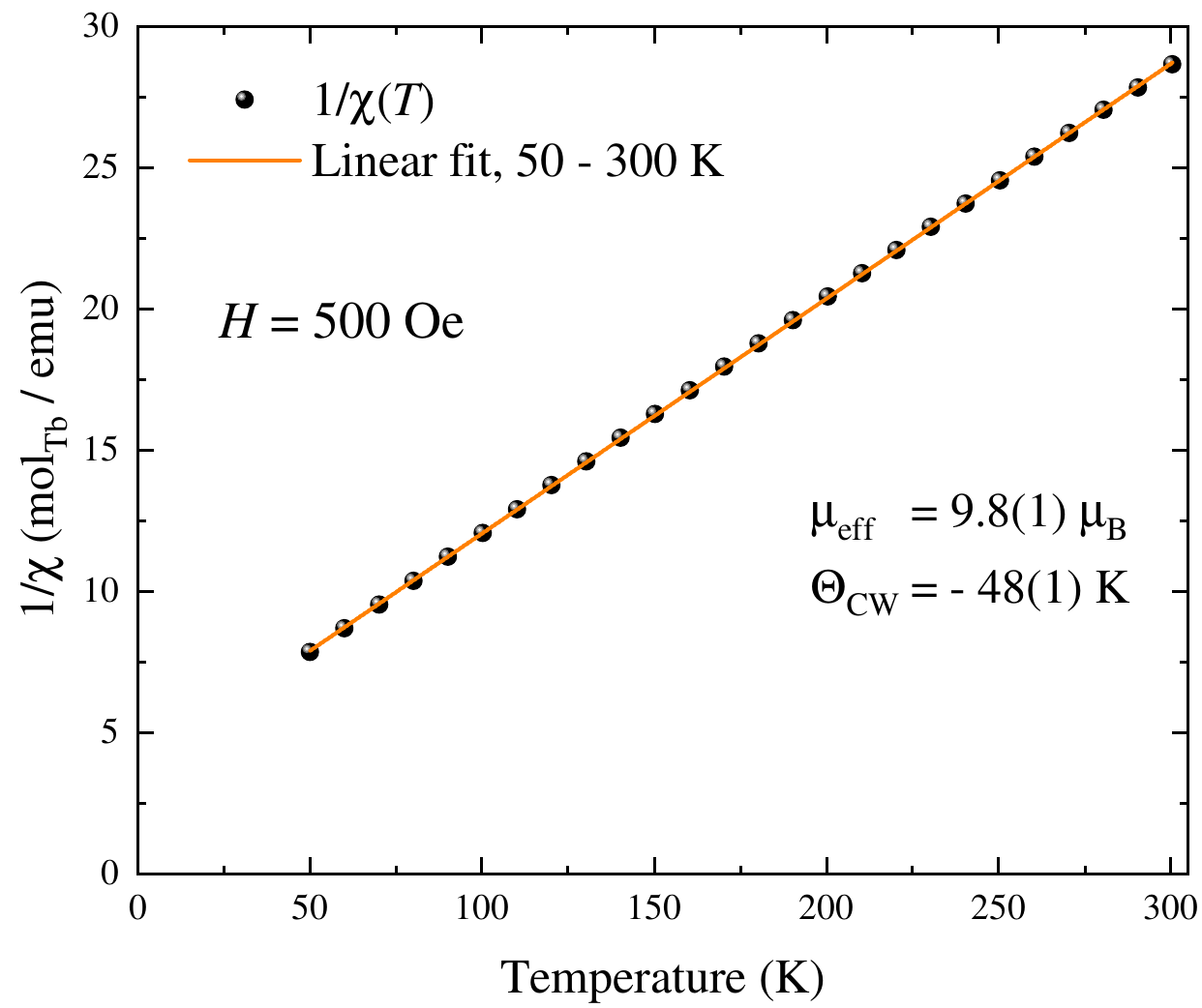}
\caption{Temperature dependence of the inverse magnetic susceptibility, $1/\chi(T)$ measured on a single crystal sample of \sto\ in a field of 500~Oe.
The data shown are averaged over three directions of applied field, $H \parallel a$, $H \parallel b$, and $H \parallel c$.}
\label{fig:chi_avr_xtal}
\end{figure}

Figure~\ref{fig:chi_avr_xtal} shows temperature dependence of the inverse magnetic susceptibility, $1/\chi(T)$, measured on a single crystal sample of \sto.
Linear fit (solid line) is applied to the data obtained by averaging magnetic susceptibility for the field applied along the three principal directions.  

\subsection{Unpolarised neutron diffraction, zero field}
\begin{figure}[tb]
\includegraphics[width=0.99\columnwidth]{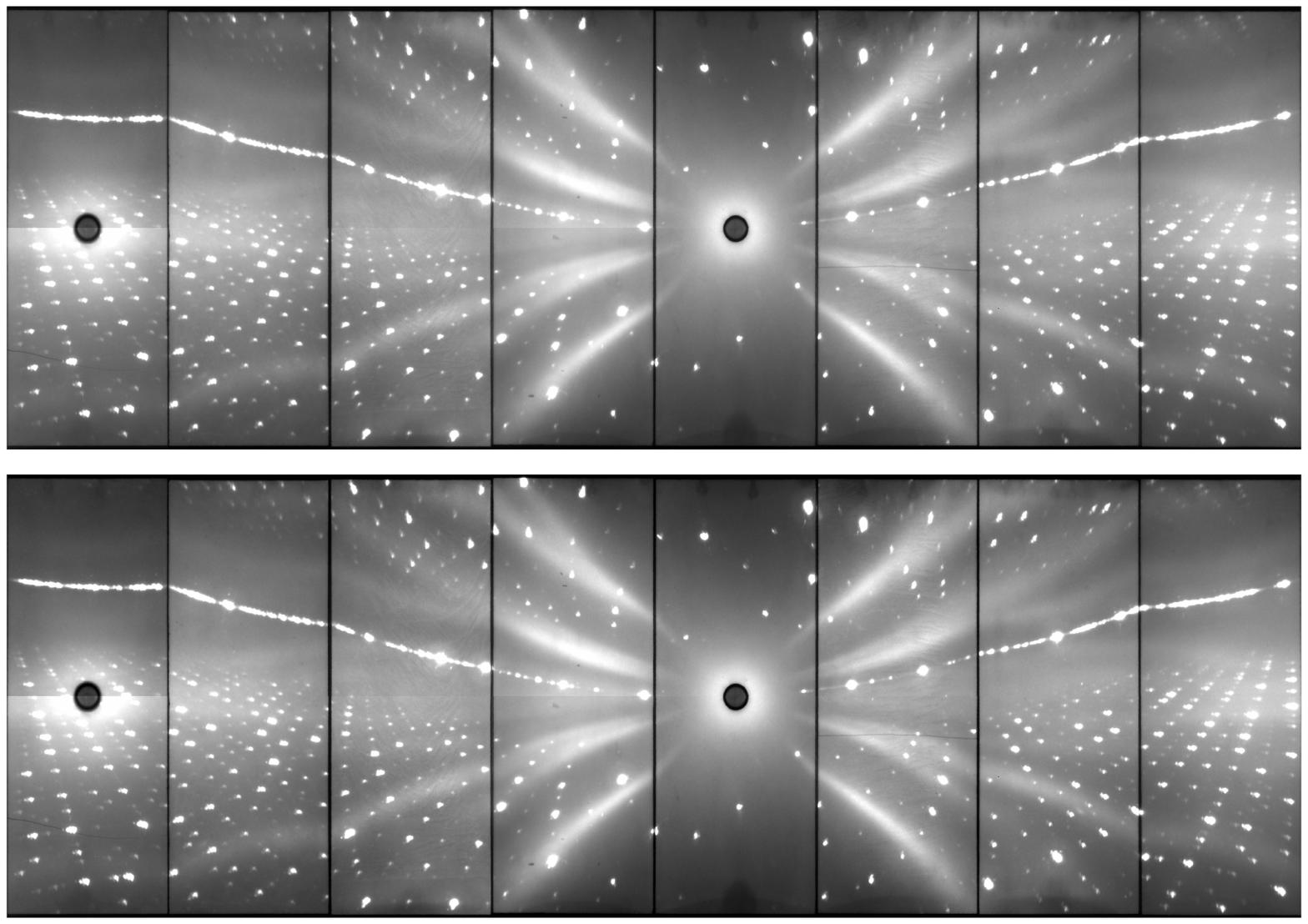}
\caption{Neutron Laue diffraction patterns of \sto\ single crystal taken on the CYCLOPS diffractometer at the ILL at (top) 2.0~K and (bottom) 10.0~K.}
\label{fig:cyclops}
\end{figure} 

Figure~\ref{fig:cyclops} shows the neutron Laue diffraction patterns of \sto\ taken at 2.0~K (top panel) and 10~K (bottom panel).
The patterns are virtually identical, and they demonstrate no discernible temperature variation.
The sample was rotated by 45~degrees compared to the orientation used in Fig.~8 of the main text.

\subsection{Polarised neutron diffraction, zero field}
\begin{figure}[tb]
\includegraphics[width=0.75\columnwidth]{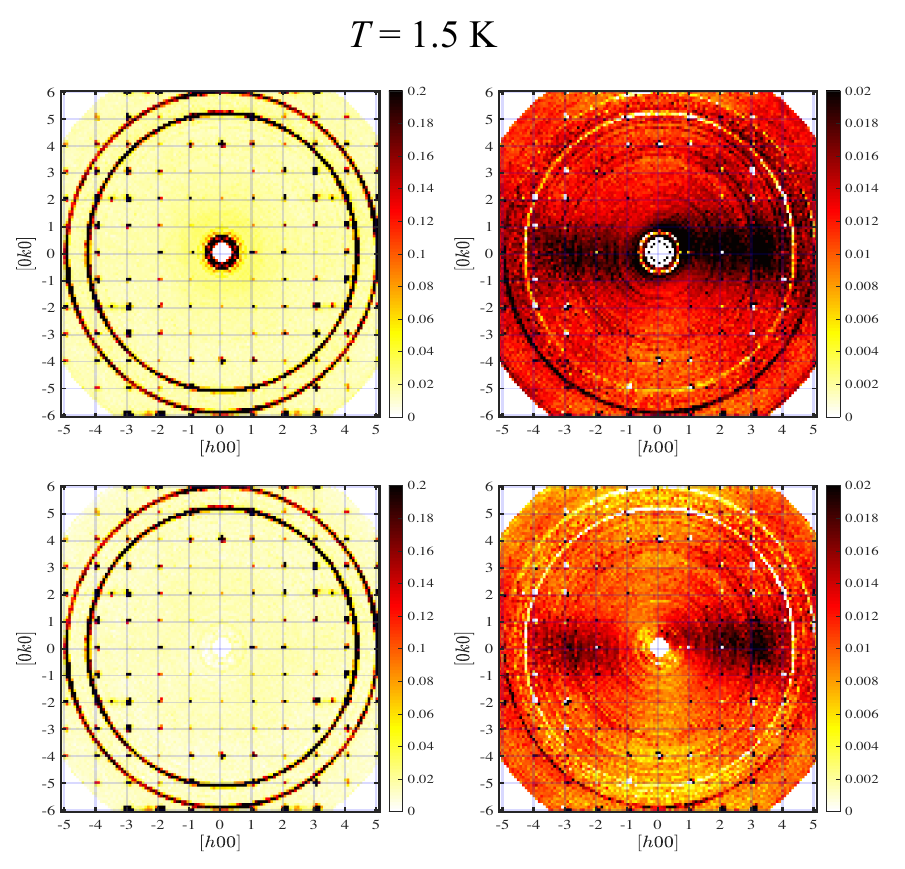}
\caption{As-collected single crystal neutron diffraction measurements on the D7 polarised neutron instrument at 1.5~K.
The intensity maps of the $(hk0)$ plane for the non-spin-flip (left) and spin-flip (right) channels are shown.
For the data shown in the lower panels, the empty-cryostat background was subtracted.}
\label{fig:D7a}
\end{figure}

\begin{figure}[tb]
\includegraphics[width=0.85\columnwidth]{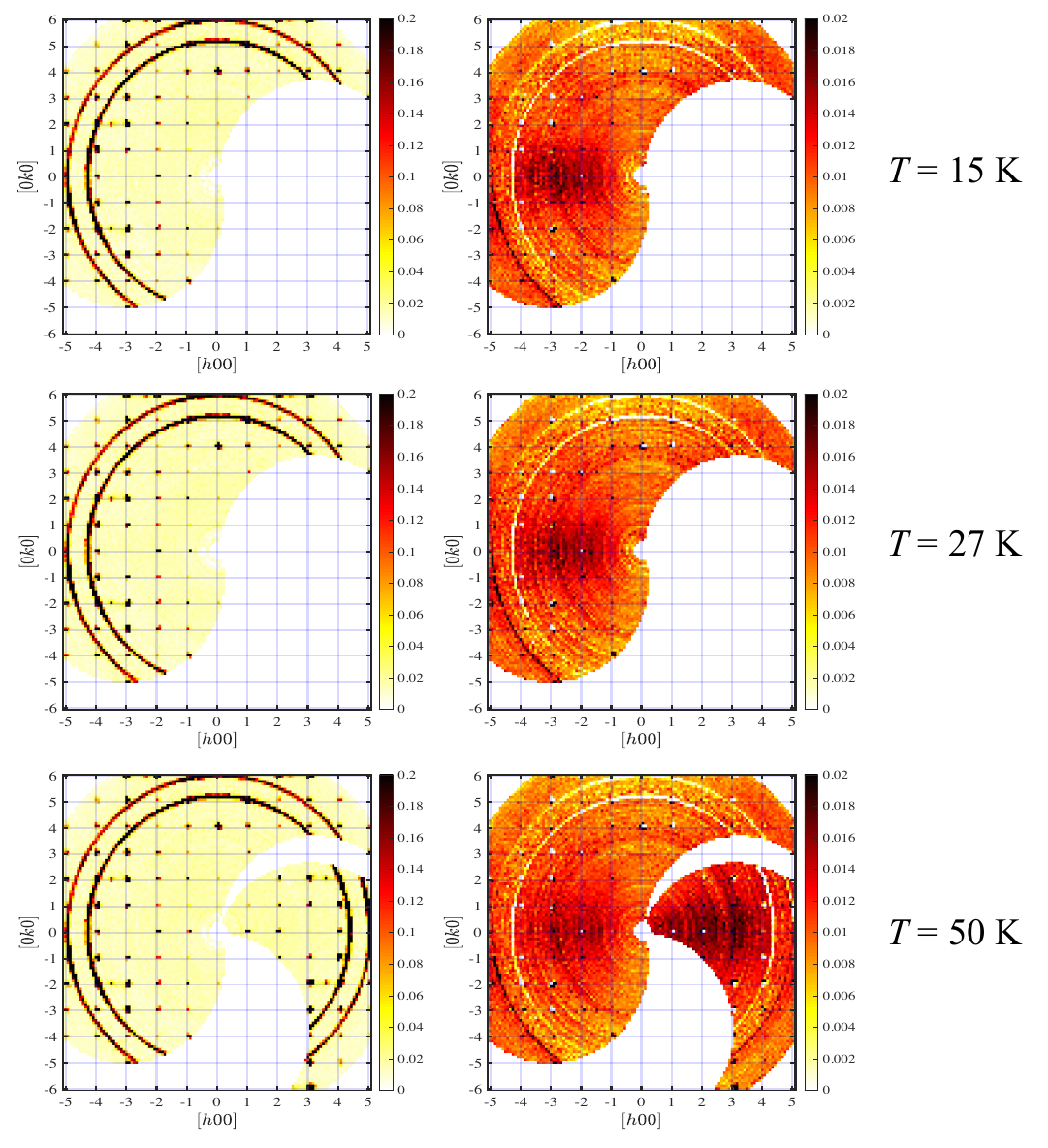}
\caption{As-collected single crystal neutron diffraction measurements on the D7 polarised neutron instrument at 15, 27 and 50~K (top to bottom  panels).
The intensity maps of the $(hk0)$ plane for the non-spin-flip (left) and spin-flip (right) channels are shown with the empty-cryostat background subtracted.}
\label{fig:D7b}
\end{figure}

The results of the single crystal neutron diffraction measurements performed on the D7 instrument with polarised neutrons are presented in Figs~\ref{fig:D7a} and~\ref{fig:D7b}.
Note that the full 360$^\circ$ maps were collected only for the base temperature of 1.5~K (as shown in Fig.~\ref{fig:D7a}), while the data sets for higher temperatures cover a smaller portion of the reciprocal space. 

\subsection{Unpolarised neutron diffraction in an applied field}
\begin{figure}[tb]
\subcaptionbox{Tb1}
{\includegraphics[width=0.33\columnwidth]{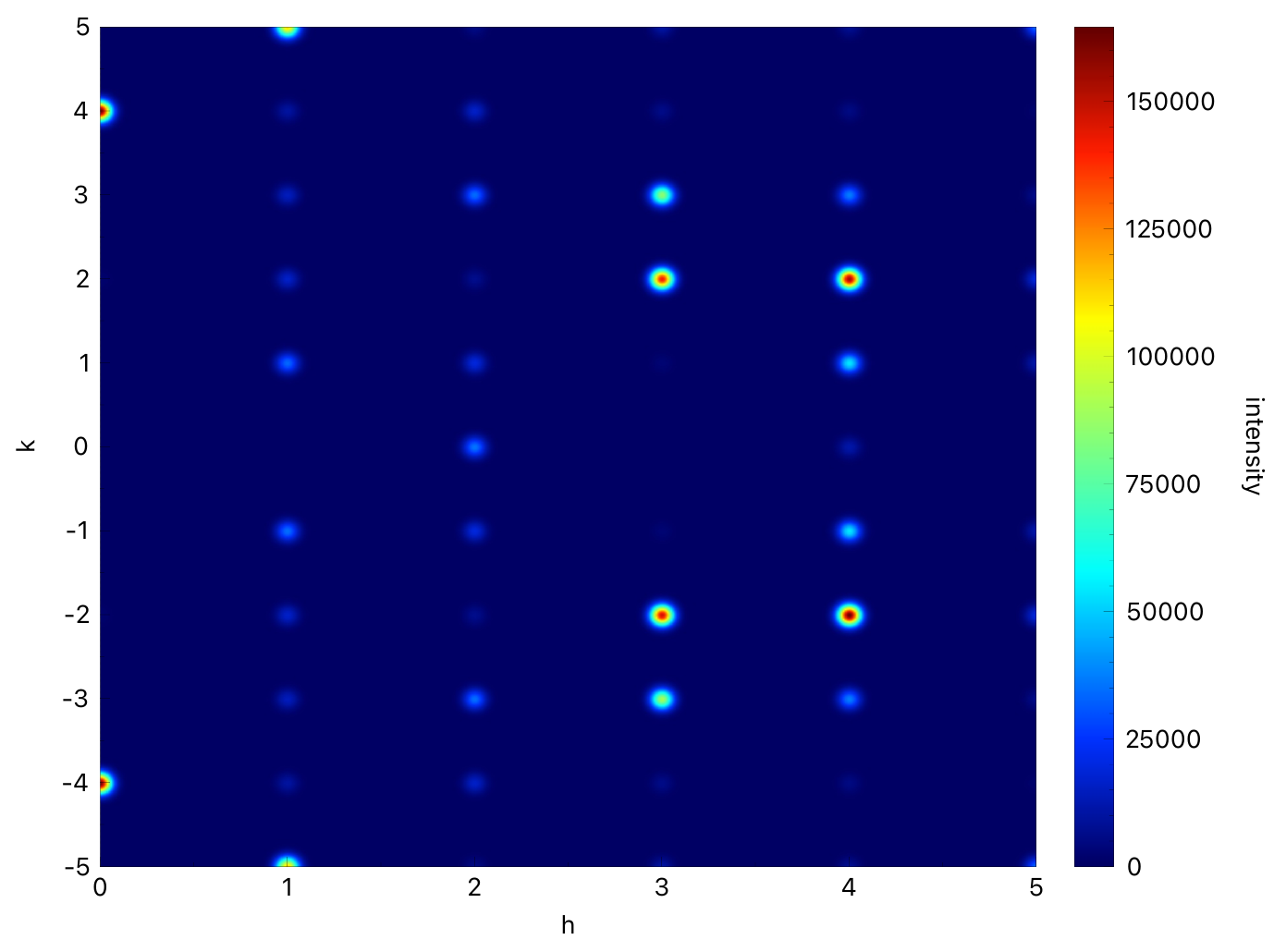}}
\subcaptionbox{Tb2}
{\includegraphics[width=0.33\columnwidth]{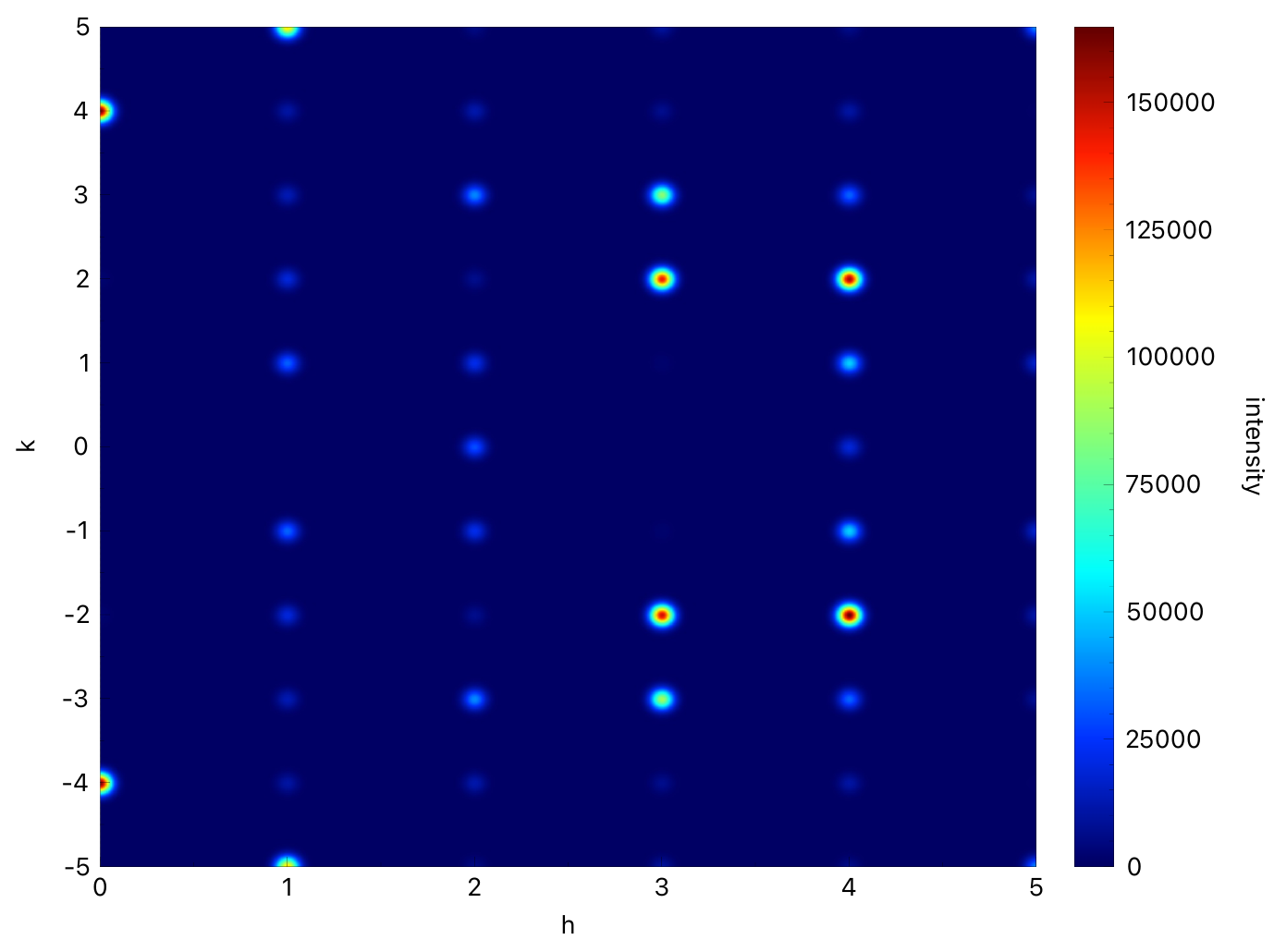}}
\subcaptionbox{Tb1 and Tb2}
{\includegraphics[width=0.33\columnwidth]{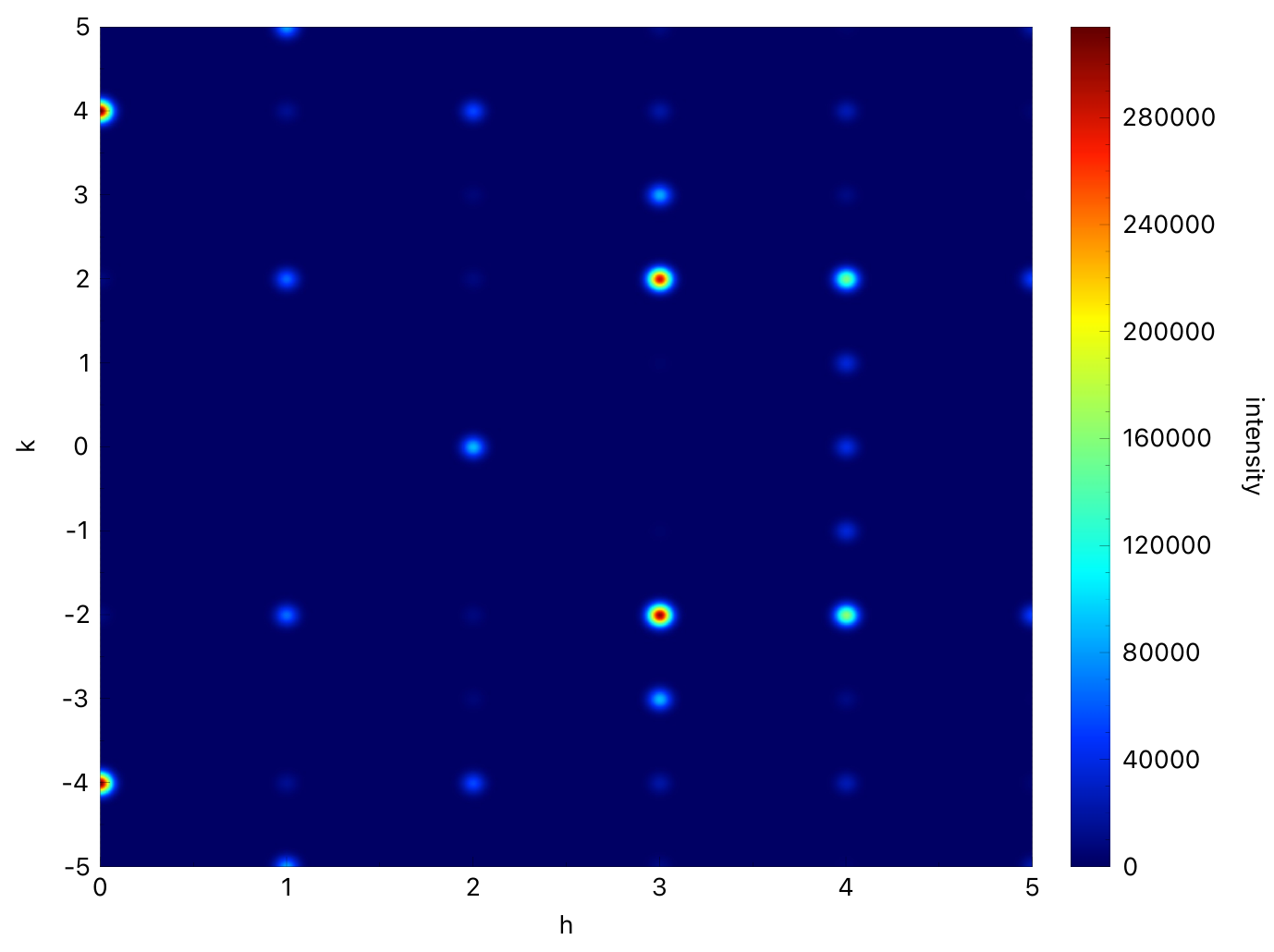}}
\caption{Calculated neutron diffraction maps for magnetic scattering from \sto\ single crystal in the $(hk0)$ scattering plane.
The equal size magnetic moments are presumed to be aligned parallel to the $c$~axis.
The calculations are the magnetic moments only on the (a) Tb1 site, (b) the Tb2 site, and (c) both sites (right panel).}
\label{fig:Mag2Pol}
\end{figure}

Figure~\ref{fig:Mag2Pol} shows the predicted neutron diffraction intensity maps for the $(hk0)$ scattering plane.
It illustrates the point made in the Section~IIID3 of the main text that the magnetic intensity maps look very similar for the models with only Tb1 or Tb2 having magnetic moment on them, but rather different if both Tb sites carry the same magnetic moment.

\onecolumngrid

\bibliography{SrLn2O4_all}